\documentclass{article}
\usepackage{epsfig}
\usepackage{cite}
\def\bom#1{{\mbox{\boldmath $#1$}}}
\def\MSbar{$\overline{{\rm MS}}$}
\def\lapprox{\lower .7ex\hbox{$\;\stackrel{\textstyle <}{\sim}\;$}}
\def\gapprox{\lower .7ex\hbox{$\;\stackrel{\textstyle >}{\sim}\;$}}

\def\e{\epsilon}
\def\d{{\rm d}}
\def\Li{\, {\rm Li}}
\def\S{\, {\rm S}}
\def\G{{\rm G}}
\def\H{{\rm H}}
\def\f{{\rm f}}
\def\g{{\rm g}}
\def\CA{C_A}

\def\NF{N_F}
\def\sab{s_{12}}
\def\sac{s_{13}}
\def\sbc{s_{23}}

\def\sabc{s_{123}}

\def\ket#1{|{#1}\rangle}
\def\bra#1{\langle{#1}|}
\def\braket#1#2{\langle #1 |#2 \rangle}
\def\cm{{\cal M}}

\def\Poles{{\cal P}oles}
\def\Finite{{\cal F}inite}
\def\Bab{{\rm Bub}((1-y-z)s_{123})}
\def\Bac{{\rm Bub}(ys_{123})}
\def\Bbc{{\rm Bub}(zs_{123})}
\def\Babc{{\rm Bub}(s_{123})}
\def\Boxyz{s_{123}{\rm Box}^6(ys_{123},zs_{123},s_{123})}
\def\Boxxz{s_{123}{\rm Box}^6((1-y-z)s_{123},zs_{123},s_{123})}

\def\boxLO{
\mbox{\parbox{2.5cm}{\hspace{0.25cm}
\begin{picture}(2,1)
\thicklines
\put(0.2,0){\vector(-1,0){0.1}}
\put(1.8,0){\vector(1,0){0.1}}
\put(1.8,1){\vector(1,0){0.1}}
\put(0.3,1){\vector(1,0){0.1}}
\put(0,0){\line(1,0){2}}
\put(0,1){\line(1,0){2}}
\put(0.5,0){\line(0,1){1}}
\put(1.5,0){\line(0,1){1}}
\put(0.15,0.2){$p_1$}
\put(1.7,0.2){$p_3$}
\put(1.7,0.7){$p_2$}
\put(-0.2,0.7){$p_{123}$}
\end{picture}
}} 
\hfill}
\newcommand{\bubbleLO}[1]{
\mbox{\parbox{2.5cm}{\hspace{0.25cm}
\begin{picture}(2,1)
\thicklines
\put(0.3,0.5){\vector(1,0){0.1}}
\put(0,0.5){\line(1,0){0.5}}
\put(1.5,0.5){\line(1,0){0.5}}
\put(1,0.5){\circle{1}}
\put(0.25,0.7){\makebox(0,0)[b]{$#1$}}
\end{picture}
}}
\hfill}
\newcommand{\bubbleNLO}[1]{
\mbox{\parbox{2.5cm}{\hspace{0.25cm}
\begin{picture}(2,1)
\thicklines
\put(0.3,0.5){\vector(1,0){0.1}}
\put(0,0.5){\line(1,0){2}}
\put(1,0.5){\circle{1}}
\put(0.25,0.7){\makebox(0,0)[b]{$#1$}}
\end{picture}
}}
\hfill}

\newcommand{\doublebubbleNLO}[1]{
\mbox{\parbox{3.5cm}{\hspace{0.25cm}
\begin{picture}(3,1)
\thicklines
\put(0.3,0.5){\vector(1,0){0.1}}
\put(0,0.5){\line(1,0){0.5}}
\put(2.5,0.5){\line(1,0){0.5}}
\put(1,0.5){\circle{1}}
\put(2,0.5){\circle{1}}
\put(0.25,0.7){\makebox(0,0)[b]{$#1$}}
\end{picture}
}}
\hfill}

\newcommand{\doublebubbleaNLO}[3]{
\mbox{\parbox{3.5cm}{\hspace{0.25cm}
\begin{picture}(3,1.3)
\thicklines
\put(0.3,0.5){\vector(1,0){0.1}}
\put(0,0.5){\line(1,0){0.5}}
\put(2.5,0.5){\vector(1,0){0.5}}
\put(1.5,1.1){\line(0,-1){0.6}}
\put(1.5,0.8){\vector(0,-1){0.1}}
\put(1,0.5){\circle{1}}
\put(2,0.5){\circle{1}}
\put(0.25,0.7){\makebox(0,0)[b]{$#1$}}
\put(1.5,1.2){\makebox(0,0)[b]{$#2$}}
\put(2.8,0.7){\makebox(0,0)[b]{$#3$}}
\end{picture}
}}
\hfill}

\newcommand{\trianglexNLO}[3]{
\mbox{\parbox{3cm}{\hspace{0.25cm}
\begin{picture}(2.5,1.4)
\thicklines
\put(0.3,0.7){\vector(1,0){0.1}}
\put(1.7,0.2){\vector(1,0){0.1}}
\put(1.7,1.2){\vector(1,0){0.1}}
\put(0,0.7){\line(1,0){0.5}}
\put(0.5,1.2){\oval(1,1)[br]}
\put(1,1.2){\line(1,0){1}}
\put(1,0.2){\line(1,0){1}}
\put(1,0.7){\circle{1}}
\put(0.25,0.9){\makebox(0,0)[b]{$#1$}}
\put(2.05,1.2){\makebox(0,0)[l]{$#2$}}
\put(2.05,0.2){\makebox(0,0)[l]{$#3$}}
\end{picture}
}}
\hfill}

\newcommand{\triangleNLO}[3]{
\mbox{\parbox{3cm}{\hspace{0.25cm}
\begin{picture}(2.5,1.4)
\thicklines
\put(0.3,0.7){\vector(1,0){0.1}}
\put(1.7,0.2){\vector(1,0){0.1}}
\put(1.7,1.2){\vector(1,0){0.1}}
\put(0,0.7){\line(1,0){0.5}}
\put(1,1.2){\line(0,-1){1}}
\put(1,1.2){\line(1,0){1}}
\put(1,0.2){\line(1,0){1}}
\put(1,0.7){\circle{1}}
\put(0.25,0.9){\makebox(0,0)[b]{$#1$}}
\put(2.05,1.2){\makebox(0,0)[l]{$#2$}}
\put(2.05,0.2){\makebox(0,0)[l]{$#3$}}
\end{picture}
}}
\hfill}

\newcommand{\trianglebNLO}[3]{
\mbox{\parbox{3cm}{\hspace{0.25cm}
\begin{picture}(2.5,1.4)
\thicklines
\put(0.3,0.7){\vector(1,0){0.1}}
\put(1.9,0.2){\vector(1,0){0.1}}
\put(1.9,1.2){\vector(1,0){0.1}}
\put(0,0.7){\line(1,0){0.5}}
\put(0.5,0.7){\line(2,1){1}}
\put(0.5,0.7){\line(2,-1){1}}
\put(1.5,1.2){\line(0,-1){1}}
\put(1.5,1.2){\line(1,0){0.5}}
\put(1.5,0.2){\line(1,0){0.5}}
\put(1.5,0.2){\line(-1,2){0.4}}
\put(0.25,0.9){\makebox(0,0)[b]{$#1$}}
\put(2.05,1.2){\makebox(0,0)[l]{$#2$}}
\put(2.05,0.2){\makebox(0,0)[l]{$#3$}}
\end{picture}
}}
\hfill}

\newcommand{\boxbubbleaNLO}[4]{
\mbox{\parbox{4cm}{\hspace{0.25cm}
\begin{picture}(3.5,1.4)
\thicklines
\put(0.7,0.2){\vector(-1,0){0.1}}
\put(2.8,0.2){\vector(1,0){0.1}}
\put(2.8,1.2){\vector(1,0){0.1}}
\put(0.8,1.2){\vector(1,0){0.1}}
\put(0.5,0.2){\line(1,0){2.5}}
\put(0.5,1.2){\line(1,0){2.5}}
\put(1,0.2){\line(0,1){1}}
\put(2,0.7){\circle{1}}
\put(0.45,1.2){\makebox(0,0)[r]{$#1$}}
\put(0.45,0.2){\makebox(0,0)[r]{$#2$}}
\put(3.05,1.2){\makebox(0,0)[l]{$#3$}}
\put(3.05,0.2){\makebox(0,0)[l]{$#4$}}
\end{picture}
}} 
\hfill}

\newcommand{\boxbubblebNLO}[4]{
\mbox{\parbox{4cm}{\hspace{0.25cm}
\begin{picture}(3.5,1.4)
\thicklines
\put(0.7,0.2){\vector(-1,0){0.1}}
\put(2.8,0.2){\vector(1,0){0.1}}
\put(2.8,1.2){\vector(1,0){0.1}}
\put(0.8,1.2){\vector(1,0){0.1}}
\put(0.5,0.2){\line(1,0){2.5}}
\put(0.5,1.2){\line(1,0){2.5}}
\put(2.5,0.2){\line(0,1){1}}
\put(1.5,0.7){\circle{1}}
\put(0.45,1.2){\makebox(0,0)[r]{$#1$}}
\put(0.45,0.2){\makebox(0,0)[r]{$#2$}}
\put(3.05,1.2){\makebox(0,0)[l]{$#3$}}
\put(3.05,0.2){\makebox(0,0)[l]{$#4$}}
\end{picture}
}} 
\hfill}

\newcommand{\boxxaNLO}[4]{
\mbox{\parbox{3.5cm}{\hspace{0.25cm}
\begin{picture}(3,1.4)
\thicklines
\put(0.7,0.2){\vector(-1,0){0.1}}
\put(2.3,0.2){\vector(1,0){0.1}}
\put(2.3,1.2){\vector(1,0){0.1}}
\put(0.8,1.2){\vector(1,0){0.1}}
\put(0.5,0.2){\line(1,0){2}}
\put(2,0.2){\line(-1,1){1}}
\put(0.5,1.2){\line(1,0){2}}
\put(1,0.2){\line(0,1){1}}
\put(2,0.2){\line(0,1){1}}
\put(0.45,1.2){\makebox(0,0)[r]{$#1$}}
\put(0.45,0.2){\makebox(0,0)[r]{$#2$}}
\put(2.55,1.2){\makebox(0,0)[l]{$#3$}}
\put(2.55,0.2){\makebox(0,0)[l]{$#4$}}
\end{picture}
}} 
\hfill}

\newcommand{\boxxbNLO}[4]{
\mbox{\parbox{3.5cm}{\hspace{0.25cm}
\begin{picture}(3,1.4)
\thicklines
\put(0.7,0.2){\vector(-1,0){0.1}}
\put(2.3,0.2){\vector(1,0){0.1}}
\put(2.3,1.2){\vector(1,0){0.1}}
\put(0.8,1.2){\vector(1,0){0.1}}
\put(0.5,0.2){\line(1,0){2}}
\put(1,0.2){\line(1,1){1}}
\put(0.5,1.2){\line(1,0){2}}
\put(1,0.2){\line(0,1){1}}
\put(2,0.2){\line(0,1){1}}
\put(0.45,1.2){\makebox(0,0)[r]{$#1$}}
\put(0.45,0.2){\makebox(0,0)[r]{$#2$}}
\put(2.55,1.2){\makebox(0,0)[l]{$#3$}}
\put(2.55,0.2){\makebox(0,0)[l]{$#4$}}
\end{picture}
}} 
\hfill}

\newcommand{\boxxbpNLO}[4]{
\mbox{\parbox{3.5cm}{\hspace{0.25cm}
\begin{picture}(3,1.4)
\thicklines
\put(0.7,0.2){\vector(-1,0){0.1}}
\put(2.3,0.2){\vector(1,0){0.1}}
\put(2.3,1.2){\vector(1,0){0.1}}
\put(0.8,1.2){\vector(1,0){0.1}}
\put(0.5,0.2){\line(1,0){2}}
\put(1.0,0.2){\line(1,2){0.5}}
\put(0.5,1.2){\line(1,0){2}}
\put(1,0.2){\line(0,1){1}}
\put(2,0.2){\line(0,1){1}}
\put(0.45,1.2){\makebox(0,0)[r]{$#1$}}
\put(0.45,0.2){\makebox(0,0)[r]{$#2$}}
\put(2.55,1.2){\makebox(0,0)[l]{$#3$}}
\put(2.55,0.2){\makebox(0,0)[l]{$#4$}}
\end{picture}
}} 
\hfill}

\newcommand{\doubleboxNLO}[4]{
\mbox{\parbox{3.5cm}{\hspace{0.25cm}
\begin{picture}(3,1.4)
\thicklines
\put(0.7,0.2){\vector(-1,0){0.1}}
\put(2.3,0.2){\vector(1,0){0.1}}
\put(2.3,1.2){\vector(1,0){0.1}}
\put(0.8,1.2){\vector(1,0){0.1}}
\put(0.5,0.2){\line(1,0){2}}
\put(1,0.2){\line(0,1){1}}
\put(0.5,1.2){\line(1,0){2}}
\put(1.5,0.2){\line(0,1){1}}
\put(2,0.2){\line(0,1){1}}
\put(0.45,1.2){\makebox(0,0)[r]{$#1$}}
\put(0.45,0.2){\makebox(0,0)[r]{$#2$}}
\put(2.55,1.2){\makebox(0,0)[l]{$#3$}}
\put(2.55,0.2){\makebox(0,0)[l]{$#4$}}
\end{picture}
}}
\hfill}

\newcommand{\doubleboxNLOtwo}[4]{
\mbox{\parbox{3.5cm}{\hspace{0.25cm}
\begin{picture}(3,1.4)
\thicklines
\put(0.7,0.2){\vector(-1,0){0.1}}
\put(2.3,0.2){\vector(1,0){0.1}}
\put(2.3,1.2){\vector(1,0){0.1}}
\put(0.8,1.2){\vector(1,0){0.1}}
\put(0.5,0.2){\line(1,0){2}}
\put(1,0.2){\line(0,1){1}}
\put(0.5,1.2){\line(1,0){2}}
\put(1.5,0.2){\line(0,1){1}}
\put(2,0.2){\line(0,1){1}}
\put(0.45,1.2){\makebox(0,0)[r]{$#1$}}
\put(0.45,0.2){\makebox(0,0)[r]{$#2$}}
\put(2.55,1.2){\makebox(0,0)[l]{$#3$}}
\put(2.55,0.2){\makebox(0,0)[l]{$#4$}}
\put(1.075,1.0){\makebox(0,0)[l]{$_{(2)}$}}
\end{picture}
}}
\hfill}

\newcommand{\boxxbdotNLO}[4]{
\mbox{\parbox{3.5cm}{\hspace{0.25cm}
\begin{picture}(3,1.4)
\thicklines
\put(0.7,0.2){\vector(-1,0){0.1}}
\put(2.3,0.2){\vector(1,0){0.1}}
\put(2.3,1.2){\vector(1,0){0.1}}
\put(0.8,1.2){\vector(1,0){0.1}}
\put(0.5,0.2){\line(1,0){2}}
\put(1,0.2){\line(1,1){1}}
\put(1.5,0.7){\circle*{0.2}}
\put(0.5,1.2){\line(1,0){2}}
\put(1,0.2){\line(0,1){1}}
\put(2,0.2){\line(0,1){1}}
\put(0.45,1.2){\makebox(0,0)[r]{$#1$}}
\put(0.45,0.2){\makebox(0,0)[r]{$#2$}}
\put(2.55,1.2){\makebox(0,0)[l]{$#3$}}
\put(2.55,0.2){\makebox(0,0)[l]{$#4$}}
\end{picture}
}} 
\hfill}

\newcommand{\boxbubblebexNLO}[4]{
\mbox{\parbox{4cm}{\hspace{0.25cm}
\begin{picture}(3.5,1.4)
\thicklines
\put(0.7,0.2){\vector(-1,0){0.1}}
\put(2.8,0.2){\vector(1,0){0.1}}
\put(2.8,1.2){\vector(1,0){0.1}}
\put(0.8,1.2){\vector(1,0){0.1}}
\put(0.5,0.2){\line(1,0){2.5}}
\put(0.5,1.2){\line(1,0){0.5}}
\put(1.5,1.2){\line(1,0){1.5}}
\put(2.5,0.2){\line(0,1){1}}
\put(1.5,1.2){\line(0,-1){1}}
\put(1.25,1.2){\circle{0.5}}
\put(0.45,1.2){\makebox(0,0)[r]{$#1$}}
\put(0.45,0.2){\makebox(0,0)[r]{$#2$}}
\put(3.05,1.2){\makebox(0,0)[l]{$#3$}}
\put(3.05,0.2){\makebox(0,0)[l]{$#4$}}
\end{picture}
}} 
\hfill}

\newcommand{\trianglecrossNLO}[3]{
\mbox{\parbox{3cm}{\hspace{0.25cm}
\begin{picture}(2.5,1.4)
\thicklines
\put(0.3,0.7){\vector(1,0){0.1}}
\put(1.9,0.2){\vector(1,0){0.1}}
\put(1.9,1.2){\vector(1,0){0.1}}
\put(0,0.7){\line(1,0){0.5}}
\put(0.5,0.7){\line(1,1){0.5}}
\put(0.5,0.7){\line(1,-1){0.5}}
\put(1.5,1.2){\line(-1,-2){0.5}}
\put(1.0,1.2){\line(1,-2){0.18}}
\put(1.5,0.2){\line(-1,2){0.18}}
\put(1,1.2){\line(1,0){1}}
\put(1,0.2){\line(1,0){1}}
\put(0.25,0.9){\makebox(0,0)[b]{$#1$}}
\put(2.05,1.2){\makebox(0,0)[l]{$#2$}}
\put(2.05,0.2){\makebox(0,0)[l]{$#3$}}
\end{picture}
}}
\hfill}

\newcommand{\boxxbmcrossxNLO}[4]{
\mbox{\parbox{3.5cm}{\hspace{0.25cm}
\begin{picture}(3,1.4)
\thicklines
\put(0.7,0.2){\vector(-1,0){0.1}}
\put(2.3,0.2){\vector(1,0){0.1}}
\put(2.3,1.2){\vector(1,0){0.1}}
\put(0.8,1.2){\vector(1,0){0.1}}
\put(0.5,0.2){\line(1,0){2}}
\put(1.5,0.2){\line(1,2){0.2}}
\put(2,1.2){\line(-1,-2){0.2}}
\put(0.5,1.2){\line(1,0){2}}
\put(1,1.2){\line(1,-2){0.5}}
\put(1.5,1.2){\line(1,-2){0.5}}
\put(0.45,1.2){\makebox(0,0)[r]{$#1$}}
\put(0.45,0.2){\makebox(0,0)[r]{$#2$}}
\put(2.55,1.2){\makebox(0,0)[l]{$#3$}}
\put(2.55,0.2){\makebox(0,0)[l]{$#4$}}
\end{picture}
}} 
\hfill}

\newcommand{\boxxbmcrossxNLOtwo}[4]{
\mbox{\parbox{3.5cm}{\hspace{0.25cm}
\begin{picture}(3,1.4)
\thicklines
\put(0.7,0.2){\vector(-1,0){0.1}}
\put(2.3,0.2){\vector(1,0){0.1}}
\put(2.3,1.2){\vector(1,0){0.1}}
\put(0.8,1.2){\vector(1,0){0.1}}
\put(0.5,0.2){\line(1,0){2}}
\put(1.5,0.2){\line(1,2){0.2}}
\put(2,1.2){\line(-1,-2){0.2}}
\put(0.5,1.2){\line(1,0){2}}
\put(1,1.2){\line(1,-2){0.5}}
\put(1.5,1.2){\line(1,-2){0.5}}
\put(0.45,1.2){\makebox(0,0)[r]{$#1$}}
\put(0.45,0.2){\makebox(0,0)[r]{$#2$}}
\put(2.55,1.2){\makebox(0,0)[l]{$#3$}}
\put(2.55,0.2){\makebox(0,0)[l]{$#4$}}
\put(1.15,1.0){\makebox(0,0)[l]{$_{(2)}$}}
\end{picture}
}} 
\hfill}

\newcommand{\doublecrossNLO}[4]{
\mbox{\parbox{3.5cm}{\hspace{0.25cm}
\begin{picture}(3,1.4)
\thicklines
\put(0.7,0.2){\vector(-1,0){0.1}}
\put(2.3,0.2){\vector(1,0){0.1}}
\put(2.3,1.2){\vector(1,0){0.1}}
\put(0.8,1.2){\vector(1,0){0.1}}
\put(0.5,0.2){\line(1,0){2}}
\put(1.5,0.2){\line(1,2){0.2}}
\put(2,1.2){\line(-1,-2){0.2}}
\put(0.5,1.2){\line(1,0){2}}
\put(1,0.2){\line(0,1){1}}
\put(1.5,1.2){\line(1,-2){0.5}}
\put(0.45,1.2){\makebox(0,0)[r]{$#1$}}
\put(0.45,0.2){\makebox(0,0)[r]{$#2$}}
\put(2.55,1.2){\makebox(0,0)[l]{$#3$}}
\put(2.55,0.2){\makebox(0,0)[l]{$#4$}}
\end{picture}
}} 
\hfill}

\newcommand{\doublecrossNLOtwo}[4]{
\mbox{\parbox{3.5cm}{\hspace{0.25cm}
\begin{picture}(3,1.4)
\thicklines
\put(0.7,0.2){\vector(-1,0){0.1}}
\put(2.3,0.2){\vector(1,0){0.1}}
\put(2.3,1.2){\vector(1,0){0.1}}
\put(0.8,1.2){\vector(1,0){0.1}}
\put(0.5,0.2){\line(1,0){2}}
\put(1.5,0.2){\line(1,2){0.2}}
\put(2,1.2){\line(-1,-2){0.2}}
\put(0.5,1.2){\line(1,0){2}}
\put(1,0.2){\line(0,1){1}}
\put(1.5,1.2){\line(1,-2){0.5}}
\put(0.45,1.2){\makebox(0,0)[r]{$#1$}}
\put(0.45,0.2){\makebox(0,0)[r]{$#2$}}
\put(2.55,1.2){\makebox(0,0)[l]{$#3$}}
\put(2.55,0.2){\makebox(0,0)[l]{$#4$}}
\put(1.075,1.0){\makebox(0,0)[l]{$_{(2)}$}}
\end{picture}
}} 
\hfill}

\newcommand{\doublecrossxNLO}[4]{
\mbox{\parbox{3.5cm}{\hspace{0.25cm}
\begin{picture}(3,1.4)
\thicklines
\put(0.7,0.2){\vector(-1,0){0.1}}
\put(2.3,0.2){\vector(1,0){0.1}}
\put(2.3,1.2){\vector(1,0){0.1}}
\put(0.8,1.2){\vector(1,0){0.1}}
\put(0.5,0.2){\line(1,0){2}}
\put(1,0.2){\line(1,2){0.2}}
\put(1.5,1.2){\line(-1,-2){0.2}}
\put(0.5,1.2){\line(1,0){2}}
\put(2,0.2){\line(0,1){1}}
\put(1,1.2){\line(1,-2){0.5}}
\put(0.45,1.2){\makebox(0,0)[r]{$#1$}}
\put(0.45,0.2){\makebox(0,0)[r]{$#2$}}
\put(2.55,1.2){\makebox(0,0)[l]{$#3$}}
\put(2.55,0.2){\makebox(0,0)[l]{$#4$}}
\end{picture}
}} 
\hfill}

\newcommand{\doublecrossxNLOtwo}[4]{
\mbox{\parbox{3.5cm}{\hspace{0.25cm}
\begin{picture}(3,1.4)
\thicklines
\put(0.7,0.2){\vector(-1,0){0.1}}
\put(2.3,0.2){\vector(1,0){0.1}}
\put(2.3,1.2){\vector(1,0){0.1}}
\put(0.8,1.2){\vector(1,0){0.1}}
\put(0.5,0.2){\line(1,0){2}}
\put(1,0.2){\line(1,2){0.2}}
\put(1.5,1.2){\line(-1,-2){0.2}}
\put(0.5,1.2){\line(1,0){2}}
\put(2,0.2){\line(0,1){1}}
\put(1,1.2){\line(1,-2){0.5}}
\put(0.45,1.2){\makebox(0,0)[r]{$#1$}}
\put(0.45,0.2){\makebox(0,0)[r]{$#2$}}
\put(2.55,1.2){\makebox(0,0)[l]{$#3$}}
\put(2.55,0.2){\makebox(0,0)[l]{$#4$}}
\put(1.575,1.0){\makebox(0,0)[l]{$_{(3)}$}}
\end{picture}
}} 
\hfill}

\begin{document}
\unitlength1cm
\begin{titlepage}
\vspace*{-1cm}
\begin{flushright}
IPPP/01/59\\
DCTP/01/118\\
CERN-TH/2001-348\\
hep-ph/0112081\\
December 2001 
\end{flushright}                                
\vskip 3.5cm

\begin{center}
\boldmath
{\Large\bf The Two-Loop QCD Matrix Element for $e^+e^- \to 3$~Jets}
\unboldmath
\vskip 1.cm
{\large L.W.~Garland}$^a$, {\large T.~Gehrmann}$^b$, 
{\large E.W.N.~Glover}$^a$, {\large A.~Koukoutsakis}$^a$
and {\large E.~Remiddi}$^c$ 
\vskip .7cm
{\it $^a$ Department of Physics, University of Durham, Durham DH1 3LE, England}
\vskip .4cm
{\it $^b$ Theory Division, CERN, CH-1211 Geneva 23, Switzerland}
\vskip .4cm
{\it $^c$ Dipartimento di Fisica,
    Universit\`{a} di Bologna and INFN, Sezione di 
    Bologna,  I-40126 Bologna, Italy} 
\end{center}
\vskip 2.6cm

\begin{abstract}
We compute the ${\cal O}(\alpha_s^3)$ virtual QCD corrections to the 
$\gamma^*\to q\bar q g$ matrix element arising from the 
interference of the two-loop with the tree-level amplitude
and from the self-interference of the one-loop amplitude. 
The calculation is performed by reducing all loop integrals 
appearing in the two-loop amplitude to a small set of known master integrals. 
Infrared and ultraviolet divergences are both regularized using 
conventional dimensional regularization, and the ultraviolet 
renormalization is performed in the $\overline{{\rm MS}}$ scheme. 
The infrared pole structure of the matrix elements agrees with the
prediction made by Catani using an infrared factorization formula. The 
analytic result for the finite terms of both matrix elements is expressed 
in terms of one- and two-dimensional harmonic polylogarithms.

\end{abstract}

\vfill
\end{titlepage}                                                                
\newpage

\renewcommand{\theequation}{\mbox{\arabic{section}.\arabic{equation}}}

\section{Introduction}
\setcounter{equation}{0}

Among jet observables, the three-jet production rate in  electron--positron
annihilation plays an outstanding role.  The initial experimental observation
of three-jet events at PETRA~\cite{petra}, in agreement with the theoretical
prediction~\cite{ellis}, provided  first evidence for the gluon, and thus 
strong support for the theory of Quantum Chromodynamics (QCD). Subsequently 
the three-jet rate  and related event shape observables were used
for the precise determination  of the QCD coupling constant $\alpha_s$
(see~\cite{bethke} for a review).  Especially at LEP, three-jet observables
were measured to a very high  precision and the error on the extraction of
$\alpha_s$ from these  data is dominated by the uncertainty inherent in the
theoretical  next-to-leading order (NLO)  calculation~\cite{ert1,ert2,kn,gg,cs}
of the jet observables.  The planned  TESLA~\cite{tesla} linear $e^+e^-$
collider will allow precision  QCD studies at even higher energies than at
LEP. Given the projected  luminosity of TESLA, one again expects the
experimental errors to  be well below the uncertainty of the NLO calculation. 

Related to  $e^+e^- \to 3$~jets  by crossing symmetry  are $(2+1)$-jet
production in  deep inelastic $ep$ scattering and vector-boson-plus-jet
production at hadron  colliders. The experimental data from HERA on $ep \to
(2+1)$~jets and related event shape observables have already reached a level
of  precision demanding predictions beyond the present NLO accuracy;  a
further improvement on these data is expected soon from the HERA  high
luminosity programme. Similarly,  vector-boson production at large transverse
momentum is a classic test of QCD in hadron--hadron collisions and demands the
theoretical prediction to be as precise as possible. In this case, it is also
an important  background in searches for new physics at the Tevatron and the
LHC.

Besides its  phenomenological importance,  the three-jet rate has also served
as a theoretical testing ground for the development of new techniques for
higher order calculations in QCD: both the subtraction~\cite{ert1} and the
phase-space slicing~\cite{ert2} methods for the extraction of infrared
singularities from  NLO real radiation  processes were
developed in the context of the first three-jet calculations. The systematic
formulation of  phase-space  slicing~\cite{gg} as  well as the dipole
subtraction~\cite{cs} method were also first demonstrated for three-jet
observables, before being applied to other processes. It is very likely that
similar techniques at higher orders will first be developed in the context of
jet production in $e^+e^-$ annihilation, which in contrast to hadron--hadron
collisions or electron--proton scattering does not pose the additional
difficulty of the regularization of  initial state singularities.

The calculation of  next-to-next-to-leading order (NNLO), i.e.\  ${\cal
O}(\alpha_s^3)$, corrections to the three-jet rate in $e^+e^-$ annihilation 
has been considered as a highly important project for a long 
time~\cite{kunszt}. In terms of matrix elements, it requires the computation
of three contributions: the tree level $\gamma^* \to 5$~partons
amplitudes~\cite{tree5p1,tree5p2,tree5p3}, the one-loop corrections to the $\gamma^* \to
4$~partons amplitudes~\cite{onel4p1,onel4p2,onel4p3,onel4p4},  and the two-loop
(as well as the one-loop  times one-loop) corrections to the  $\gamma^* \to
3$~partons matrix  elements. While the  former two contributions have been 
known for some time already, the two-loop amplitudes have presented  an
obstacle that prevented further progress on this calculation up to now. 

This calculation has now become tractable owing to
various technical  developments over the last two years. In particular, the
systematic application of  integration-by-parts~\cite{hv,chet1,chet2} and 
Lorentz invariance~\cite{gr} identities allowed the large number of
Feynman integrals appearing  in two-loop  four-point matrix elements 
to be reduced to a small
number of so-called master integrals. The use of these techniques already 
allowed the calculation of  two-loop QED and QCD corrections to many $2\to 2$
scattering processes with massless on-shell external 
particles~\cite{m1,m2,m3,m4,m5,m6}, which require master integrals
corresponding to  massless four-point functions with all legs 
on-shell~\cite{onshell1,onshell2,onshell3,onshell4,onshell5,onshell6}. The
master integrals relevant in the context of the present work are massless
four-point functions with three legs on-shell and one leg off-shell. The
complete set of these integrals was computed in~\cite{mi}, earlier partial
results had been presented in~\cite{num,smirnov}.  

The master integrals in~\cite{mi} are expressed in terms of  two-dimensional
harmonic polylogarithms (2dHPLs).  The 2dHPLs are an extension of the harmonic 
polylogarithms (HPLs) of~\cite{hpl}. All HPLs and 2dHPLs that appear in the 
divergent parts of the planar master integrals have weight  $\leq 3$ and can
be  related to the more commonly known Nielsen  generalized 
polylogarithms~\cite{nielsen,bit} of suitable arguments.  The functions of
weight 4 appearing in the finite  parts of the master integrals can all be
represented, by the very  definition, as one-dimensional  integrals over
2dHPLs of weight 3, hence of Nielsen's  generalized
polylogarithms of suitable arguments according to the above  remark. A table
with all relations  is included in the appendix of~\cite{mi}. Numerical
routines providing an evaluation of   the HPLs~\cite{grnum1} and
2dHPLs~\cite{grnum2} are available.

In this paper, we present the ${\cal O}(\alpha_s^3)$ corrections to  the
$\gamma^*\to q\bar q g$ matrix element.
At this order, two combinations of amplitudes 
contribute: the  interference of
two-loop and tree amplitudes and the self-interference of the 
one-loop amplitude. We work in conventional  dimensional
regularization~\cite{dreg1,dreg2,hv}, with $d=4-2\e$
space-time dimensions, where all external
particles are $d$-dimensional.  Ultraviolet renormalization is performed in the
$\overline{{\rm MS}}$ scheme.   The infrared pole structure of the two-loop
corrections to the  $\gamma^*\to q\bar q g$ matrix element was predicted by 
Catani~\cite{catani}, using an infrared factorization formula.  We confirm
Catani's prediction with our explicit calculation, and  we use the formalism
introduced in~\cite{catani} to present the  infrared poles and the finite
parts  of the  $\gamma^*\to q\bar q g$ matrix elements in a compact form. 

The paper is structured as follows. In Section~\ref{sec:note}, we define  the
notation and kinematics used in the paper. Section~\ref{sec:method} explains
the method we used to reduce the two-loop diagrams to  master integrals. The
result for the two-loop QCD contribution to  the $\gamma^*\to q\bar q g$ matrix
element, decomposed into  infrared-divergent and infrared-finite parts
according  to the prescription derived in~\cite{catani}, is given in 
Section~\ref{sec:me}. Finally, Section~\ref{sec:conc} contains the  conclusions
and an outlook on future steps needed for the  completion of a full NNLO
calculation of three-jet production in  $e^+e^-$ annihilation.

\section{Notation}
\label{sec:note}
\setcounter{equation}{0}
We consider the decay of a virtual photon into a quark--antiquark--gluon 
system:
\begin{equation}
\gamma^* (q) \longrightarrow q(p_1) + \bar q (p_2) + g(p_3)\; .
\end{equation}
The kinematics of this process is fully described by the invariants
\begin{equation}
\sab = (p_1+p_2)^2\;, \qquad \sac = (p_1+p_3)^2\;, \qquad 
\sbc = (p_2+p_3)^2\;,
\end{equation}
which fulfil
\begin{equation}
q^2 = \sab + \sac + \sbc \equiv \sabc \; .
\end{equation}
It is convenient to define the dimensionless invariants
\begin{equation}
x = \sab/\sabc\;, \qquad y = \sac/\sabc\;, \qquad z = \sbc/\sabc\;,
\end{equation}
with $x+y+z=1$.

Our calculation is performed in conventional dimensional regularization
\cite{dreg1,dreg2,hv} with 
$d=4-2\e$, and all external particle states are taken to be 
$d$-dimensional. Renormalization of ultraviolet divergences is 
performed in the $\overline{{\rm MS}}$ scheme. The renormalized 
amplitude can be written as
\begin{equation}
|{\cal M}\rangle = \sqrt{4\pi\alpha}e_q \sqrt{4\pi\alpha_s} \left[
|{\cal M}^{(0)}\rangle 
+ \left(\frac{\alpha_s}{2\pi}\right) |{\cal M}^{(1)}\rangle 
+ \left(\frac{\alpha_s}{2\pi}\right)^2 |{\cal M}^{(2)}\rangle 
+ {\cal O}(\alpha_s^3) \right] \;,
\label{eq:renorme}
\end{equation}
where $\alpha$ denotes the electromagnetic coupling constant, 
$e_q$ the quark charge,
$\alpha_s$ the QCD coupling constant at the renormalization scale $\mu$, 
and the $|{\cal M}^{(i)}\rangle$ are the $i$-loop contributions to the 
renormalized amplitude. They are vectors in colour space. 

The squared amplitude, summed over spins, colours and quark flavours, 
is denoted by
\begin{equation}
\langle{\cal M}|{\cal M}\rangle = \sum |{\cal M}(\gamma^* \to q\bar q g)|^2 
= {\cal T} (x,y,z)\; .
\end{equation}
The perturbative expansion of ${\cal T} (x,y,z)$ at renormalization scale 
$\mu^2 = q^2 = s_{123}$ reads:
\begin{eqnarray}
{\cal T} (x,y,z) &=& 16\pi^2\alpha\sum_q e_q^2 \alpha_s(q^2)\Bigg[
{\cal T}^{(2)} (x,y,z) + 
\left(\frac{\alpha_s(q^2)}{2\pi}\right){\cal T}^{(4)} (x,y,z) \nonumber \\
&& \hspace{1.3cm}
+ \left(\frac{\alpha_s(q^2)}{2\pi}\right)^2{\cal T}^{(6)} (x,y,z) 
 + {\cal O}(\alpha_s^3(q^2)) \Bigg] \;,
\end{eqnarray}
where 
\begin{eqnarray}
\label{eq:T2}
{\cal T}^{(2)} (x,y,z) &=& \langle{\cal M}^{(0)}|{\cal M}^{(0)}\rangle 
= 4 V (1-\e)\left[ (1-\e)\left(\frac{y}{z}+\frac{z}{y}\right) 
+\frac{2(1-y-z)-2\e yz}{yz}\right]\;,\\
\label{eq:T4}
{\cal T}^{(4)} (x,y,z) &=& 
\langle{\cal M}^{(0)}|{\cal M}^{(1)}\rangle +
\langle{\cal M}^{(1)}|{\cal M}^{(0)}\rangle \; ,\\
\label{eq:T6}
{\cal T}^{(6)} (x,y,z) &=& 
\langle{\cal M}^{(1)}|{\cal M}^{(1)}\rangle +
\langle{\cal M}^{(0)}|{\cal M}^{(2)}\rangle +
\langle{\cal M}^{(2)}|{\cal M}^{(0)}\rangle \;,
\end{eqnarray}
where $V=N^2-1$, with $N$ the number of colours. 
${\cal T}^{(4)} (x,y,z)$ was first derived in~\cite{ert1,ert2}; 
we quote an explicit expression for it in Section~\ref{sec:irf}.
In the following, 
we present the contribution to ${\cal T}^{(6)} (x,y,z)$ from the interference 
of two-loop and tree diagrams
\begin{equation}
{\cal T}^{(6,[2\times 0])} (x,y,z) = 
\langle{\cal M}^{(0)}|{\cal M}^{(2)}\rangle +
\langle{\cal M}^{(2)}|{\cal M}^{(0)}\rangle \;,
\end{equation}
as well as the one-loop self-interference 
\begin{equation}
{\cal T}^{(6,[1\times 1])} (x,y,z) = 
\langle{\cal M}^{(1)}|{\cal M}^{(1)}\rangle \;.
\end{equation}
At the same order in $\alpha_s$, one finds also a contribution to 
three-jet final states from the self-interference of the 
$\gamma^*\to ggg$ amplitude. The matrix element for this process does 
not contain infrared or ultraviolet divergences; it was computed long ago and 
can be found in~\cite{vg1,vg2}.

For the remainder of this paper we will set the renormalization scale
$\mu^2 = q^2$. 
The full scale dependence of the perturbative expansion is given by
\begin{eqnarray}
\label{eq:rge}
\lefteqn{
{\cal T} (x,y,z) = 16\pi^2\alpha\sum_q e_q^2 \alpha_s(\mu^2)\bigg\{
{\cal T}^{(2)} (x,y,z) }\nonumber \\
&&+ \left(\frac{\alpha_s(\mu^2)}{2\pi}\right)
\left[
{\cal T}^{(4)} (x,y,z) 
+b_0 {\cal T}^{(2)} (x,y,z)\ln\left({\mu^2\over q^2}\right)
\right]
\nonumber \\
&&+ \left(\frac{\alpha_s(\mu^2)}{2\pi}\right)^2
\bigg[
{\cal T}^{(6)} (x,y,z) 
+\biggl(2b_0{\cal T}^{(4)} (x,y,z)+b_1{\cal T}^{(2)}(x,y,z) \biggr) 
\ln\left({\mu^2\over q^2}\right)  
\nonumber\\
&&\hspace{3cm}
+b_0^2 {\cal T}^{(2)} (x,y,z)
\ln^2\left({\mu^2\over q^2}\right)\bigg]
 + {\cal O}(\alpha_s^3) \bigg\}.
\end{eqnarray}

\section{Method}
\label{sec:method}
\setcounter{equation}{0}

The Feynman diagrams  contributing to the $i$-loop 
amplitude $|{\cal M}^{(i)}\rangle$ ($i=0,1,2$) were all generated using 
QGRAF~\cite{qgraf}. There are two diagrams at tree-level, 13 diagrams 
at one loop and 229 diagrams at two loops. 
 We then project $|{\cal M}^{(2)}\rangle$ by  $\bra{\cm^{(0)}}$
and $|{\cal M}^{(1)}\rangle$ by $\bra{\cm^{(1)}}$,
and perform the summation over colours and spins using the computer algebra
programs MAPLE~\cite{maple}, FORM2~\cite{form2} and FORM3~\cite{form3}.   
When summing over the polarizations of the external 
gluon and off-shell photon,  we use the Feynman
gauge:
\begin{equation}
\sum_{{\rm spins}} \epsilon_{i}^{\mu}\epsilon_{i}^{\nu *} = 
-  g^{\mu \nu}. 
\end{equation}
This is valid because the gluon always couples to a 
conserved fermionic  current, which selects only the  physical degrees of
polarization. The use of an axial polarization sum  to project out the
transverse polarizations (as applied  in~\cite{m3,m4}) is therefore not
needed. 

The one-loop self-interference contribution ${\cal T}^{(6,[1\times 1])}$ is 
computed by reducing all tensorial loop integrals according to the standard 
Passarino--Veltman procedure~\cite{pv} to scalar one-loop 
two-point, three-point and four-point integrals. It has been known for a 
long time that those three-point integrals 
can be further reduced to linear combinations of two-point integrals 
using integration-by-parts identities. After this reduction, 
${\cal T}^{(6,[1\times 1])}$ is expressed as a bilinear 
combination of only two integrals:
the one-loop box and the one-loop bubble, which are listed in 
Appendix~\ref{app:master_int}. 

The computation of ${\cal T}^{(6,[2\times 0])}$ 
is by far less straightforward.
It is explained in detail in the following two subsections. 

\subsection{Master Integrals}

The integrals appearing in the individual  two-loop
diagrams contain up to seven
propagators  in the denominator, and up to four irreducible scalar products in
the  numerator (i.e.\ scalar products which can not be expressed 
as linear combinations of the occurring propagators). 
Each integral is classified by the number of different 
denominators $t$, the total number of denominators $r$ and the number of
irreducible scalar products $s$. The set of $t$ different denominators  defines
the topology of the diagram. 

Using the reduction procedure described in Section~\ref{sec:red} below,
all of the two-loop Feynman diagrams can be reduced to a basis set of
{\em master} integrals. Owing to the presence of the extra scale, there are
considerably more master topologies than in the on-shell case.   Altogether
there are 14 planar topologies and 5 non-planar topologies requiring a total of
24 master integrals, as five topologies possess two master integrals 
(see below). For each topology, we take one of the master integrals 
(or the master integral, when it is unique) to be equal to the full 
scalar amplitude with the first power on all propagators.
The simpler master integrals are the single
scale integrals~\cite{kl1,kl2,kl3}, which  
can be written in terms of $\Gamma$ functions,
\begin{eqnarray*}
{\rm Sunrise}(s_{12}) &=& \bubbleNLO{p_{12}}\;,\\
{\rm Glass}(s_{12}) &=&\doublebubbleNLO{p_{12}}\;,\\
{\rm Dart}_1(s_{12})&=& \triangleNLO{p_{12}}{p_1}{p_2}\;,
\end{eqnarray*}
as well as the more complicated
$$
{\rm Xtri}_1(s_{12}) =\trianglecrossNLO{p_{12}}{p_1}{p_2}\;.
$$
The two-scale master integrals can be written in terms of $\Gamma$ functions,
$$
{\rm Tglass}(s_{12},s_{123}) =\doublebubbleaNLO{p_{12}}{p_3}{p_{123}}\;,
$$
or as generalized polylogarithms or
one-dimensional
harmonic polylogarithms~(see e.g.~\cite{mi}),
\begin{eqnarray*}
{\rm Dart}_2(s_{12},s_{123}) &=& \triangleNLO{p_{123}}{p_{12}}{p_3}\;,\\
{\rm Dart}_2(s_{123},s_{23}) &=&\trianglexNLO{p_{123}}{p_1}{p_{23}}\;,\\
{\rm Plane}(s_{12},s_{123}) &=& \trianglebNLO{p_{123}}{p_{12}}{p_{3}}\;,\\
{\rm Xtri}_2(s_{123},s_{12}) &=&\trianglecrossNLO{p_{123}}{p_{12}}{p_3}\;.
\end{eqnarray*}
The three-scale master integrals can be written in terms of two-dimensional
harmonic polylogarithms~\cite{mi}.   There are the planar graphs,
\begin{eqnarray*}
{\rm Abox}_1(s_{23},s_{13},s_{123}) &=& \boxbubbleaNLO{p_{123}}{p_1}{p_2}{p_3}
\;,\\
{\rm Abox}_2(s_{23},s_{13},s_{123}) &=& \boxbubblebNLO{p_{123}}{p_1}{p_2}{p_3}
\;,\\
{\rm Cbox}_1(s_{23},s_{13},s_{123}) &=& \boxxaNLO{p_{123}}{p_1}{p_2}{p_3}\;,\\
{\rm Cbox}_2(s_{23},s_{13},s_{123}) &=& \boxxbNLO{p_{123}}{p_1}{p_2}{p_3}\;,\\
{\rm Tbox}_1(s_{23},s_{13},s_{123}) &=& \boxxbpNLO{p_{123}}{p_1}{p_2}{p_3}\;,\\
{\rm Pbox}_1(s_{23},s_{13},s_{123}) &=& \doubleboxNLO{p_{123}}{p_1}{p_2}{p_3}
\;,\\
{\rm Bbox}(s_{23},s_{13},s_{123})&=& \boxbubblebexNLO{p_{123}}{p_1}{p_2}{p_3}
\;,
\end{eqnarray*}
and the non-planar graphs,
\begin{eqnarray*}
{\rm Ebox}_1(s_{12},s_{13},s_{123}) &=& 
\boxxbmcrossxNLO{p_{123}}{p_3}{p_2}{p_1}\; ,\\
{\rm Xbmo}_1(s_{12},s_{13},s_{123}) &=& 
\doublecrossNLO{p_{123}}{p_3}{p_2}{p_1}\;,\\
{\rm Xbmi}_1(s_{12},s_{13},s_{123}) &=& 
\doublecrossxNLO{p_{123}}{p_3}{p_2}{p_1}\;.\\
\end{eqnarray*}
It turns out that for the ${\rm Cbox}_2$, ${\rm Pbox}$, ${\rm Ebox}$, ${\rm
Xbmo}$ and ${\rm Xbmi}$ topologies, a second master integral is required 
for expressing all the occurring integrals. 
One is free to choose the precise form of these additional master 
integrals in terms
of additional powers of a propagator or as a tensor integral, provided 
the second master integral is 
not directly related by LI and IBP identities to the first one. As 
in~\cite{mi} we choose as second master integral
\begin{eqnarray*}
{\rm Cbox}_{2A}(s_{23},s_{13},s_{123}) &=& 
\boxxbdotNLO{p_{123}}{p_1}{p_2}{p_3}\;,\\
{\rm Pbox}_2(s_{23},s_{13},s_{123}) &=& 
\doubleboxNLOtwo{p_{123}}{p_1}{p_2}{p_3}\;,\\
{\rm Ebox}_2(s_{12},s_{13},s_{123}) &=& 
\boxxbmcrossxNLOtwo{p_{123}}{p_3}{p_2}{p_1}\;,\\
{\rm Xbmo}_2(s_{12},s_{13},s_{123}) &=& 
\doublecrossNLOtwo{p_{123}}{p_3}{p_2}{p_1}\;,\\
{\rm Xbmi}_2(s_{12},s_{13},s_{123}) &=& 
\doublecrossxNLOtwo{p_{123}}{p_3}{p_2}{p_1}\;.
\end{eqnarray*}

\subsection{Reduction to master integrals}
\label{sec:red}

The reduction of the two-loop diagrams to master integrals using 
integration-by-parts (IBP)~\cite{hv,chet1,chet2} and Lorentz invariance
(LI)~\cite{gr} identities was  performed with two independent programs, which
allowed  a comparison  diagram by diagram. 

One program was based on the reduction algorithm already described 
in~\cite{gr}. In this program, reduction equations are derived  by applying the
differentiation operators of the IBP and LI  identities to the following
integrands:  $t=7, r=7,s\leq 4$; $3 \leq t\leq 6 ,t \leq  r \leq 7, s \leq 3$
(note that it is not necessary to  include $s=4$ integrands with $t<7$). As a
result, one  obtains a linear system of equations, which contains, besides
integrals  of the class $t,r,s$ and simpler integrals,  also more complicated
integrals of the classes $t,r+1,s$ and $t,r+1,s+1$.  The number of equations
does, however, exceed the number of integrals,  thus yielding an 
apparently overconstrained
system of equations. By solving this  system of equations using the computer
algebra programs MAPLE~\cite{maple} and FORM2~\cite{form2}, one expresses 
integrals of a given  class $t,r,s$  as linear combinations of simpler
integrals. The irreducible integrals, which are left over after this procedure,
are the master integrals. Compared to the  algorithm used in~\cite{gr}, one
major modification was made.  For integrals  corresponding to 
those topologies which
do not possess an independent
master integral (i.e.\ can be fully reduced to integrals
with one less propagator
by the IBP and LI identities), it is straightforward  to obtain a
symbolic solution  of the IBP and LI identities. This solution  takes the form
of a single  equation decreasing the power on one or more of the propagators 
(see~\cite{nigelpentabox} for an example). These symbolic reduction  equations
for all non-master topologies were derived automatically  using a
FORM2~\cite{form2} program. The use  of these symbolic  identities did 
lead to a
considerable speed-up of the  reduction of 
the amplitudes of such diagrams to master integrals. 

The second approach derived the reduction equations using the IBP and LI
identities for three auxiliary diagrams --- one planar and two non-planar.  
These auxiliary diagrams are obtained
by mapping the nine possible dot-products involving the three independent
external momenta ($p_1,~p_2$ and $p_3$) and the two
loop momenta as propagators.
Scalar products in the numerator are replaced 
by propagators raised to negative powers.  
For illustration, the planar auxiliary diagram is shown in Fig.~1, 
with each
propagator $i$ raised to the arbitrary power $\nu_i$.
An advantage of this method is that the 10 IBP and 3 LI identities are written
in terms of arbitrary $\nu_i$ and are valid for all subtopologies of the
auxiliary diagram ($\nu_i \to 0$) and all tensor integrals ($\nu_i < 0$).
Typically we encounter subtopologies with up to 7 propagators ($t=7$) and 
tensor integrals up to fourth rank ($s=4$).  For a given value of  $t$, and
assignment of propagators, we generate equations using tensor integrals with $s
\leq 4$ as seeds for the generic IBP and LI equations. These seed integrals are
those that appear in the actual calculation of Feynman diagrams.  For example,
for $t=7$,  there are 15 seed integrals yielding 195 equations.   The equations
are stored and solved  one by one  according to a ranking similar to that
proposed by Laporta~\cite{laporta} using
MAPLE~\cite{maple} and FORM3~\cite{form3}. At each stage, more complicated integrals
are decomposed in terms of simpler integrals of the same topology and pinched
integrals.   As in the first method, the remaining irreducible integrals are
master integrals.  Of course, the system of equations contains many more
complicated integrals (with, for example, $r > 7$ or $s > 4$), which cannot be
reduced to master integrals with this approach.   However, they are not needed
for the evaluation of the Feynman diagrams and can be eliminated from the
system of equations. 
\begin{figure}[tb]
   \begin{center}  
   \epsfysize=4.0cm
   \epsffile{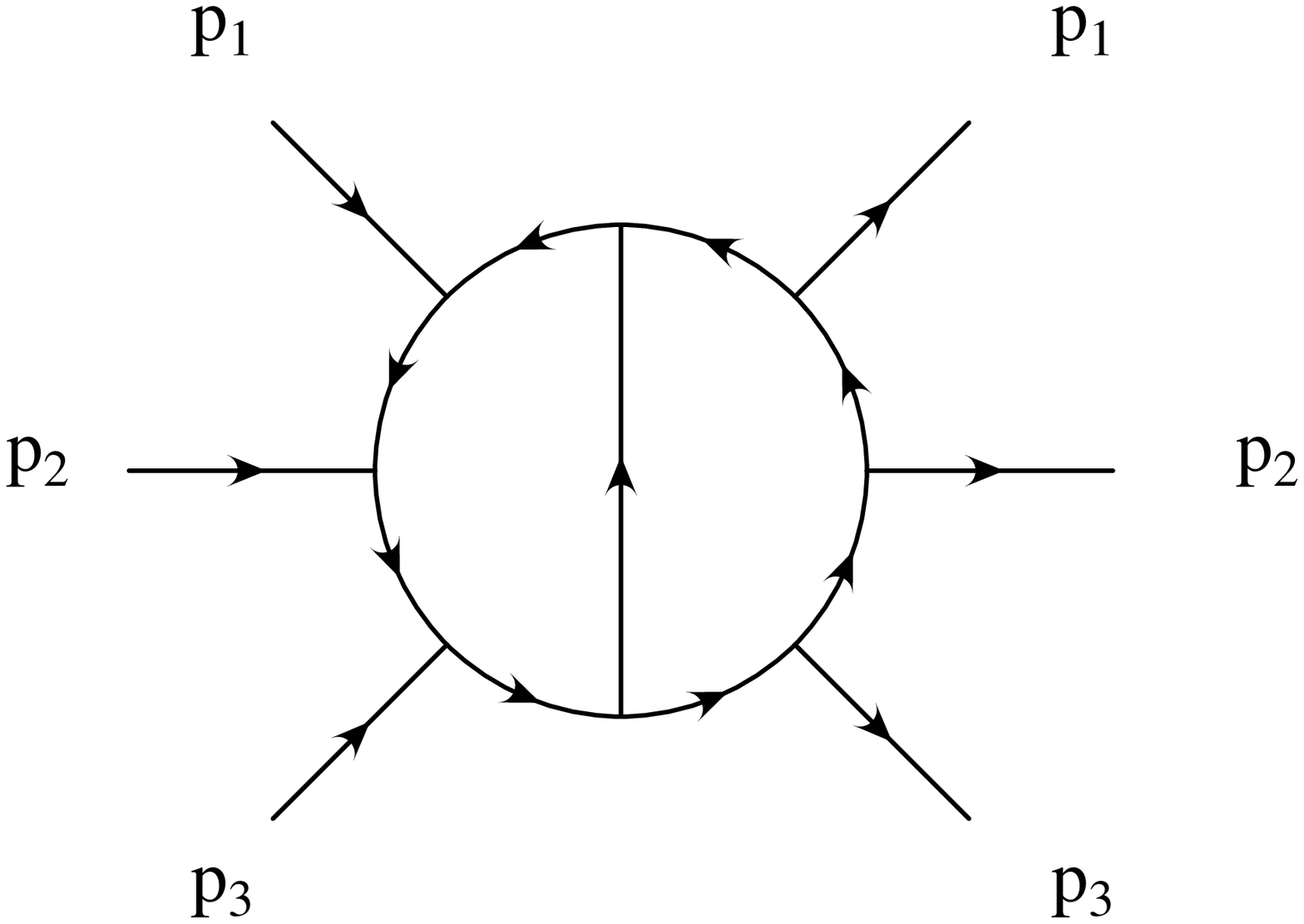}
   \hspace{1cm}
   \epsfysize=4.0cm
   \epsffile{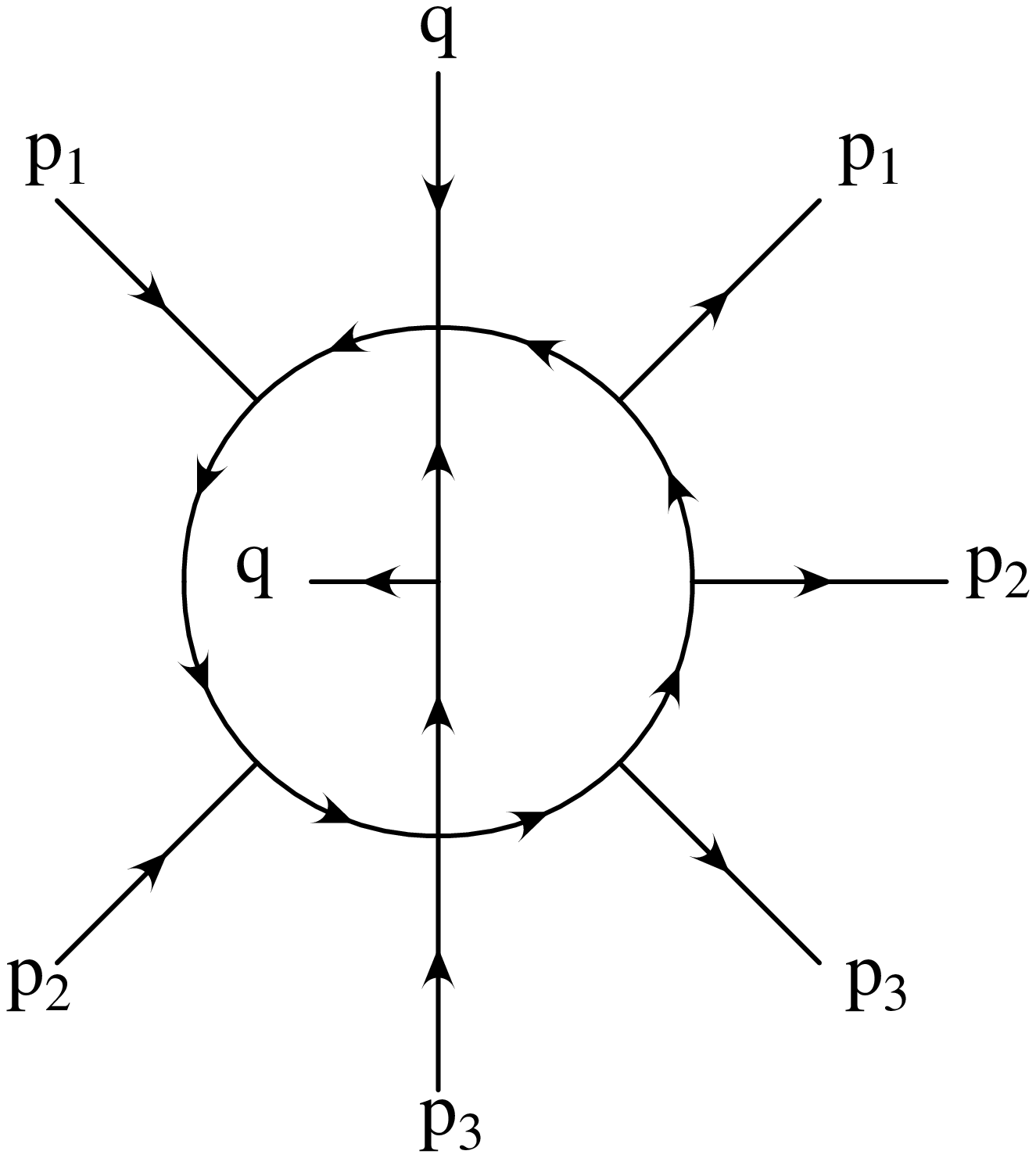}
   \hspace{1cm}
   \epsfysize=4.0cm
   \epsffile{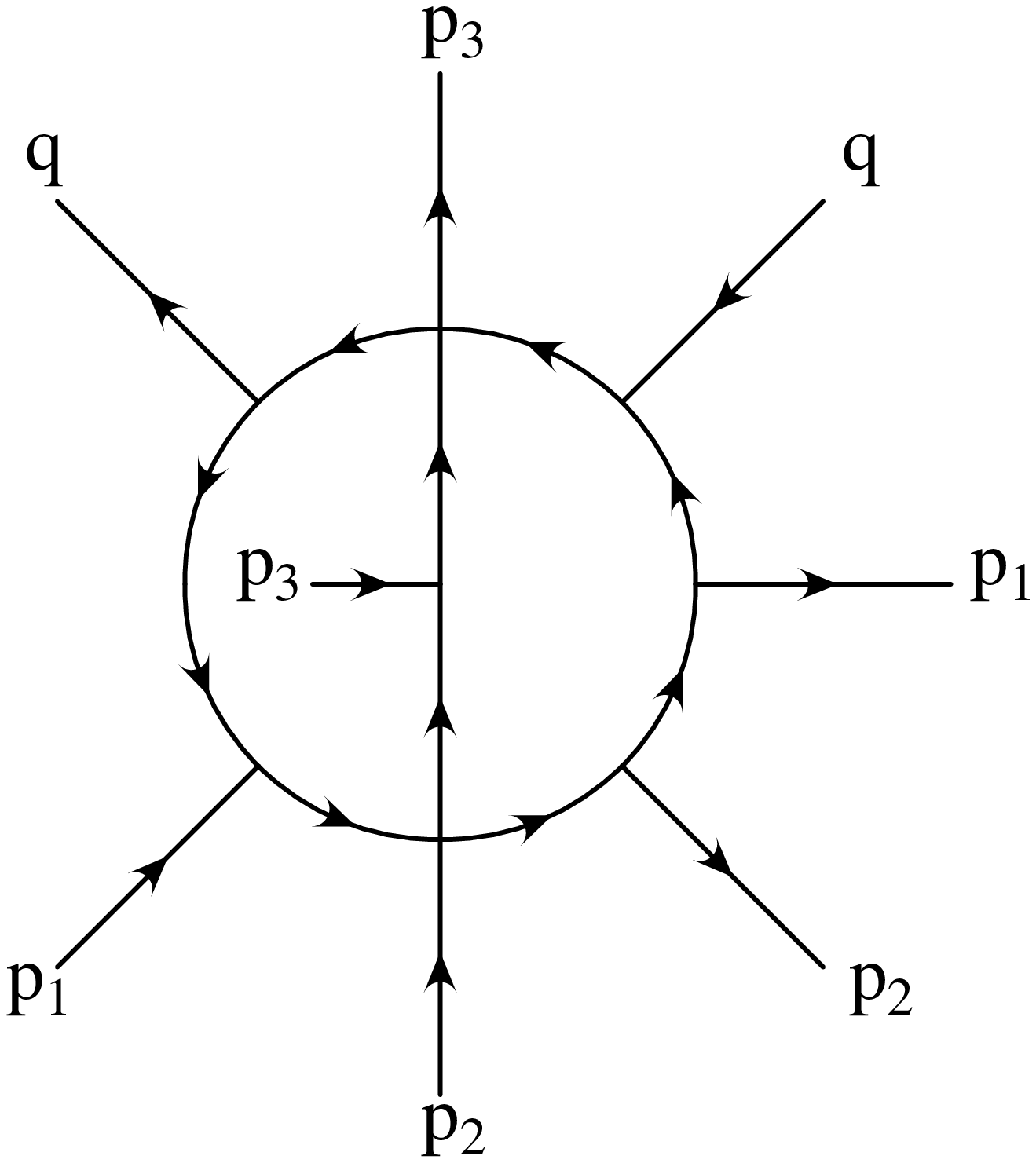}
\end{center}
\caption{The planar and two non-planar auxiliary diagrams.  All nine possible dot-products involving
the loop momenta and the external particle momenta are mapped onto the nine
propagators.}
\end{figure}

In both methods, one observes that  on the face of it, the planar and
non-planar graphs appear to be of equal difficulty --- requiring the same depth of
tensor reduction with  the same numbers of seed integrals and the same number
of IBP and LI equations. However the complexity of the non-planar graphs is
first revealed in the number of terms in each equation and then in the work
done to eliminate the unwanted integrals from the system of equations.

Both methods have their advantages. The major advantage of  the first method is
that the reduction equations for each topology  containing master integrals
have to be evaluated only once, while  the second method requires their
repeated evaluation each time  a topology appears as subtopology of the
auxiliary diagram. As a consequence,  the first method needed less computer
time for the derivation of  all required reduction equations. An advantage of
the second method  is that it needs less human intervention, since 
mapping of subtopologies to master topologies is not necessary,
 and the 
generalization to more complicated graphs with additional energy scales
requires only the modification of the 13 IBP and LI equations.   The solution
of the system and identification of master integrals is only limited by
computer resources.

\subsection{Expansion of master integrals}
The two-loop master integrals relevant to the $\gamma^* \to q\bar q g$  matrix
element are two-loop four-point functions with one leg off-shell.  These
functions were all computed in~\cite{mi} in the framework  of dimensional
regularization with $d=4-2\e$ space-time dimensions.  The results of~\cite{mi}
take the form of a Laurent series in $\e$, starting  at $\e^{-4}$, with
coefficients containing one- and two-dimensional  harmonic
polylogarithms~\cite{hpl}, which are a generalization of  Nielsen's
polylogarithms~\cite{nielsen}.  All master integrals in~\cite{mi} were given
for one particular  configuration of the external momenta. They were expressed
in  a form where the argument of the 2dHPLs was always $y$, 
while $z$ appeared 
in the index vector of the 2dHPLs; $y$ and $z$ appeared also 
as argument of the HPLs. 

Each master integral can occur in six kinematic configurations (corresponding to
the permutations of $(p_1,p_2,p_3)$). To avoid hidden  zeros (arising 
from cancellations occurring in the 
combinations of HPLs and 2dHPLs with different 
 arguments and different variables
in the index vector), we express  the master integrals for all kinematic
configurations in a unique form, which  is the same as in~\cite{mi}: the
argument of the 2dHPLs is always $y$,  the variable in their index vector is
$z$, which appears also as argument of  the HPLs. 
To obtain this unique form, we
apply the following relations  to the 2dHPLs, which are all derived following
the same lines as the  formulae for the interchange of arguments discussed in
the appendix  of~\cite{mi}:
\begin{enumerate}
\item Interchange of arguments: 
$\G(\vec m(y);z) \to \G(\vec m(z);y) + \H(\vec n;z)$ .
\item Replacement of argument variable:
$\G(\vec m(z);1-y-z) \to \G(\vec m(z);y) + \H(\vec n;z)$.
\item Replacement of index variable:
$\G(\vec m(1-y-z);y) \to \G(\vec m(z);y) + \H(\vec n;z)$.
\end{enumerate}
The relations required for the two remaining combinations 
$\G(\vec m(y);1-y-z)$ and $\G(\vec m(1-y-z);z)$ are obtained by 
applying the interchange of argument relations followed by the 
appropriate replacement relations. 

The master integrals in~\cite{mi} were derived in the kinematical  situation of
a (space-like) $1\to 3$ decay, which corresponds to the  $\gamma^* \to q\bar q
g$, such that the only  analytic continuation of  them required here is the
expansion of the overall factor in the time-like  region
\begin{equation}
{\rm Re}(-1)^{-2\e}=1-2\pi^2 \e^2+2/3\pi^4 \e^4 + {\cal O}(\e^6)\; .
\end{equation}
The analytic continuation of the master integrals to other
kinematical regions is discussed in the appendix of~\cite{mi}. 

\subsection{Ultraviolet renormalization}

The renormalization of the matrix element is carried out by replacing 
the bare coupling $\alpha_0$ with the renormalized coupling 
$\alpha_s\equiv \alpha_s(\mu^2)$,
evaluated at the renormalization scale $\mu^2$
\begin{equation}
\alpha_0\mu_0^{2\e} S_\e = \alpha_s \mu^{2\e}\left[
1- \frac{\beta_0}{\e}\left(\frac{\alpha_s}{2\pi}\right) 
+\left(\frac{\beta_0^2}{\e^2}-\frac{\beta_1}{2\e}\right)
\left(\frac{\alpha_s}{2\pi}\right)^2+{\cal O}(\alpha_s^3) \right]\; ,
\end{equation}
where
\begin{displaymath}
S_\e =(4\pi)^\e e^{-\e\gamma}\qquad \mbox{with Euler constant }
\gamma = 0.5772\ldots
\end{displaymath}
and $\mu_0^2$ is the mass parameter introduced 
in dimensional regularization~\cite{dreg1,dreg2,hv} to maintain a 
dimensionless coupling 
in the bare QCD Lagrangian density; $\beta_0$ and $\beta_1$ are the first 
two coefficients of the QCD $\beta$-function:
\begin{equation}
\beta_0 = \frac{11 \CA - 4 T_R \NF}{6},  \qquad 
\beta_1 = \frac{17 \CA^2 - 10 C_A T_R \NF- 6C_F T_R \NF}{6}\;,
\end{equation}
with the QCD colour factors
\begin{equation}
\CA = N,\qquad C_F = \frac{N^2-1}{2N},
\qquad T_R = \frac{1}{2}\; .
\end{equation}

We denote the $i$-loop contribution to the unrenormalized amplitudes by 
$|{\cal M}^{(i),{\rm un}}\rangle$, using the same normalization as 
for the decomposition of the renormalized amplitude (\ref{eq:renorme}).
The renormalized amplitudes are then obtained as
\begin{eqnarray}
|{\cal M}^{(0)}\rangle &=& |{\cal M}^{(0),{\rm un}}\rangle\; ,
 \nonumber \\
|{\cal M}^{(1)}\rangle &=& 
S_\e^{-1} |{\cal M}^{(1),{\rm un}}\rangle 
-\frac{\beta_0}{2\e} |{\cal M}^{(0),{\rm un}}\rangle\; ,  \nonumber \\
|{\cal M}^{(2)}\rangle &=& 
S_\e^{-2} |{\cal M}^{(2),{\rm un}}\rangle 
-\frac{3\beta_0}{2\e} S_\e^{-1}
|{\cal M}^{(1),{\rm un}}\rangle 
-\left(\frac{\beta_1}{4\e}-\frac{3\beta_0^2}{8\e^2}\right)
|{\cal M}^{(0),{\rm un}}\rangle\; .
\end{eqnarray}

\section{The matrix element}
\label{sec:me}
\setcounter{equation}{0}

We further decompose the renormalized one- and two-loop contributions to ${\cal
T}^{(6)}$ as a 
sum of two terms
\begin{equation}
{\cal T}^{(6, [i\times j])}(x,y,z)
 = \Poles^{(i \times j)}(x,y,z)+\Finite^{(i \times j)}(x,y,z).
\end{equation} 
$\Poles$ contains infrared singularities that will be  analytically
cancelled by those occurring in radiative processes of the
same order (ultraviolet divergences are removed by renormalization).
$\Finite$ is the renormalized remainder, which is finite as $\epsilon \to 0$.
In this section we first give explicit expressions for the infrared pole
structure using the procedure advocated by Catani~\cite{catani} and then give
the analytic results for the finite remainders.
For simplicity we set the renormalization scale $\mu^2 = s_{123}$ and 
restore the renormalization scale dependence using Eq.~(\ref{eq:rge}).

\subsection{Infrared factorization}
\label{sec:irf}

Catani~\cite{catani} has shown how to organize the 
infrared pole structure of the two-loop contributions renormalized in the 
\MSbar\ scheme in terms of the tree and renormalized one-loop amplitudes,
$\ket{\cm^{(0)}}$ and $\ket{\cm^{(1)}}$ respectively, as
\begin{eqnarray}
\label{eq:polesa}
\lefteqn{\hspace{-2cm}\Poles^{(2 \times 0)} = 
2 \Re \Biggl[  -\frac{1}{2}\bra{\cm^{(0)}} {\bom I}^{(1)}(\epsilon){\bom
I}^{(1)}(\epsilon) \ket{\cm^{(0)}}
  -\frac{\beta_0}{\epsilon}  
\,\bra{\cm^{(0)}} {\bom I}^{(1)}(\epsilon) 
\ket{\cm^{(0)}}}
 \nonumber\\
&& 
+\,  \bra{\cm^{(0)}} {\bom I}^{(1)}(\epsilon)  \ket{\cm^{(1)}}
 \nonumber\\
&& 
+
e^{-\epsilon \gamma } \frac{ \Gamma(1-2\epsilon)}{\Gamma(1-\epsilon)} 
\left(\frac{\beta_0}{\epsilon} + K\right)
\bra{\cm^{(0)}} {\bom I}^{(1)}(2\epsilon) \ket{\cm^{(0)}}\nonumber \\
&&+ \, \bra{\cm^{(0)}}{\bom H}^{(2)}(\epsilon)\ket{\cm^{(0)}} \Biggr]
\end{eqnarray}
and
\begin{equation}
\label{eq:polesb}
\Poles^{(1 \times 1)} = 
\Re \Biggl[  2 \bra{\cm^{(1)}} {\bom I}^{(1)}(\epsilon)  \ket{\cm^{(0)}}
-\bra{\cm^{(0)}} {\bom I}^{(1)\dagger}(\epsilon)
{\bom I}^{(1)}(\epsilon) \ket{\cm^{(0)}}
\, \Biggr]
\end{equation}
where the constant $K$ is
\begin{equation}
K = \left( \frac{67}{18} - \frac{\pi^2}{6} \right) \CA - 
\frac{10}{9} T_R \NF.
\end{equation}
It should be noted that, in this prescription, part of the finite 
terms in ${\cal T}^{(6,[i\times j])}$ are accounted for by the ${\cal O}(\e^0)$
expansion of $\Poles^{(i \times j)}$.

For this particular process, there is only one colour structure present at
tree level which, in terms of the gluon colour $a$ and the quark
and antiquark
colours $i$ and $j$, is simply $\bom{T}^{a}_{ij}$. Adding higher loops does not
introduce additional colour structures, and the amplitudes are therefore
vectors in a one-dimensional space.  Similarly, 
the infrared singularity operator $\bom{I}^{(1)}(\epsilon)$ is a $1 \times 1$ matrix in the colour space
and is given by
\begin{equation}
\bom{I}^{(1)}(\epsilon)
=
- \frac{e^{\epsilon\gamma}}{2\Gamma(1-\epsilon)} \Biggl[
N \left(\frac{1}{\epsilon^2}+\frac{3}{4\epsilon}+\frac{\beta_0}{2N\epsilon}\right) 
\left({\tt S}_{13}+{\tt S}_{23}\right)-\frac{1}{N}
\left(\frac{1}{\epsilon^2}+\frac{3}{2\epsilon}\right)
{\tt S}_{12}\Biggr ]\; ,\label{eq:I1}
\end{equation}
where (since we have set $\mu^2 = s_{123}$)
\begin{equation}
{\tt S}_{ij} = \left(-\frac{s_{123}}{s_{ij}}\right)^{\epsilon}.
\end{equation}
Note that on expanding ${\tt S}_{ij}$,
imaginary parts are generated, the sign of which is fixed by the small imaginary
part $+i0$ of $s_{ij}$.
Other combinations such as 
$\bra{\cm^{(0)}}\bom{I}^{(1)\dagger}(\epsilon)$  are obtained by using the hermitian conjugate
operator $\bom{I}^{(1)\dagger}(\epsilon)$, 
where the only practical change is that the sign of the
imaginary part of ${\tt S}$ is reversed.
The origin of the various terms in Eq.~(\ref{eq:I1}) is straightforward.  Each parton pair $ij$
in the event forms a radiating antenna of scale $s_{ij}$.  
Terms proportional to ${\tt S}_{ij}$ are cancelled by real radiation emitted from leg
$i$ and absorbed by leg $j$. The soft singularities ${\cal O}(1/\epsilon^2)$ are independent of
the identity of the participating partons and are universal.
However, the collinear singularities depend on the identities of the participating partons.  For
each quark we find a contribution of $3/(4\epsilon)$ and for each gluon we find a contribution
of $\beta_0/(2\epsilon)$ coming from the integral over the collinear splitting function.

Finally, the last term of Eq.~(\ref{eq:polesa}) that involves 
${\bom H}^{(2)}(\epsilon)$ 
produces only a single pole in $\epsilon$ and is given by 
\begin{equation}
\label{eq:htwo}
\bra{\cm^{(0)}}{\bom H}^{(2)}(\epsilon)\ket{\cm^{(0)}} 
=\frac{e^{\epsilon \gamma}}{4\,\epsilon\,\Gamma(1-\epsilon)} H^{(2)} 
\braket{\cm^{(0)}}{\cm^{(0)}} \;,  
\end{equation}
where the constant $H^{(2)}$ is renormalization-scheme-dependent.
As with the single pole parts of $\bom{I}^{(1)}(\epsilon)$,
the process-dependent
$H^{(2)}$ can be constructed by counting the number of
radiating partons present in the event.
In our case, there is a quark--antiquark pair and a gluon present in the final
state, so that 
\begin{equation}
H^{(2)} =  2H^{(2)}_{q}+H^{(2)}_g
\end{equation}
where in the \MSbar\ scheme
\begin{eqnarray}
H^{(2)}_g &=&  
\left(\frac{1}{2}\zeta_3+{\frac {5}{12}}+ {\frac {11\pi^2}{144}}
\right)N^2
+{\frac {5}{27}}\,\NF^2
+\left (-{\frac {{\pi }^{2}}{72}}-{\frac {89}{108}}\right ) N \NF 
-\frac{\NF}{4N}, \\
H^{(2)}_q &=&
\left({7\over 4}\zeta_3+{\frac {409}{864}}- {\frac {11\pi^2}{96}}
\right)N^2
+\left(-{1\over 4}\zeta_3-{41\over 108}-{\pi^2\over 96}\right)
+\left(-{3\over 2}\zeta_3-{3\over 32}+{\pi^2\over 8}\right){1\over
N^2}\nonumber \\
&&
+\left({\pi^2\over 48}-{25\over 216}\right){(N^2-1)N_F\over N}\;,
\end{eqnarray}
so that
\begin{eqnarray}
\label{eq:Htwo}
H^{(2)} &=&  
\left(4\zeta_3+\frac{589}{432}- \frac{11\pi^2}{72}\right)N^2
+\left(-\frac{1}{2}\zeta_3-\frac{41}{54}-\frac{\pi^2}{48} \right)
+\left(-3\zeta_3 -\frac{3}{16} + \frac{\pi^2}{4}\right) \frac{1}{N^2}\nonumber \\
&&
+\left(-\frac{19}{18}+\frac{\pi^2}{36} \right) N\NF 
+\left(-\frac{1}{54}-\frac{\pi^2}{24}\right) \frac{\NF}{N}+ \frac{5}{27} \NF^2.
\end{eqnarray}
The factors $H^{(2)}_q$ and $H^{(2)}_g$ are directly related to those found 
in gluon--gluon scattering~\cite{m4}, quark--quark scattering~\cite{m2}
and quark--gluon scattering~\cite{m3} (which each involve four partons) 
as well as in the quark form factor~\cite{kl1,qff1,qff2,qff3}.
We also note that (on purely dimensional grounds) one 
might expect terms of the 
type ${\tt S}_{ij}^2$ to be present in $H^{(2)}$.  Of course such terms are $1 + {\cal
O}(\epsilon)$ and therefore leave the pole part unchanged and only modify
the finite remainder.
At present it is not known how to systematically include these effects.
 
The renormalized interference of tree and one-loop amplitudes also appears in
Eq.~(\ref{eq:polesa}).  This can be written to all orders in $\e$ using the
relation
\begin{equation}
\braket{\cm^{(0)}}{\cm^{(1)}} =  S_\e^{-1} 
\braket{\cm^{(0),un}}{\cm^{(1),un}} - \frac{\beta_0}{2\e}
\braket{\cm^{(0),un}}{\cm^{(0),un}}\; ,
\end{equation}
where 
$$\braket{\cm^{(0),un}}{\cm^{(1),un}} = V \left(N f_1(y,z) + \frac{1}{N}
f_2(y,z)  + (y \leftrightarrow z) \right). 
$$
The functions $f_1(y,z)$ and $f_2(y,z)$ can be written in terms of the one-loop
bubble integral and the one-loop box integral in $d=6-2\e$ dimensions, Box$^6$:
\begin{eqnarray}
\label{eq:f1}
\lefteqn{f_1(y,z)=}\nonumber\\
&& {1\over yz}\bigg((-3+\e+2\e^2)\Babc+\bigg(-{4\over \e}+12-8\e\bigg)\Bac\bigg)
\nonumber \\ &&+{y\over z}\bigg(\bigg(-{2\over\e}+8-10\e+3\e^2+\e^3\bigg)\Bbc+(-3+4\e+\e^2-2\e^3)\Babc
\nonumber \\ &&\hspace{2cm}+\bigg(-{2\over \e}+8-10\e+4\e^2\bigg)\Bac\bigg)
\nonumber \\ &&+{1\over z}\bigg(\bigg({4\over \e}-12+9\e-\e^2\bigg)\Bbc+(6-2\e-4\e^2)\Babc
+\bigg({4\over \e}-12+8\e\bigg)\Bac\bigg)
\nonumber \\ &&+{y\over (1-z)^2}(1-\e)\bigg(\Bbc-\Babc\bigg)
\nonumber \\ &&+{y\over (1-z)}\bigg((3-5\e+2\e^3)\Bbc+(-3+4\e+\e^2-2\e^3)\Babc\bigg)
\nonumber \\ &&+{1\over (1-z)}(4-3\e-3\e^2-2\e^3)\bigg(\Babc-\Bbc\bigg)
\nonumber \\ &&+(4-9\e+6\e^2-\e^3)\Bbc
\nonumber \\ &&+\Boxyz(1-2\e)\bigg({1\over z}8(-1+\e)
+{y^2\over z}(-2+4\e-2\e^2)
+{y\over z}(6-8\e+2\e^2)
\nonumber \\ &&
\hspace{2cm}+z(-2+2\e-8\e^2)
+(4-3\e+3\e^2)
+{1\over yz} 2(1-\e)\bigg)\;,\\
\lefteqn{f_2(y,z)=}\nonumber \\ 
\label{eq:f2}
&&{1\over yz}\bigg((3-\e-2\e^2)\Babc+\bigg({2\over \e}-6+4\e\bigg)\Bab\bigg)
\nonumber \\ &&+{y\over z} \bigg(-\e^2(1-\e)\Bbc+(3-4\e-\e^2+2\e^3)\Babc
\nonumber \\ &&\hspace{2cm}
+\bigg({2\over \e}-8+10\e-4\e^2\bigg)\Bab\bigg)
\nonumber \\ &&+{1\over z}\bigg(\e(1-\e)\Bbc+(-6+2\e+4\e^2)\Babc+\bigg(-{4\over \e}+12-8\e\bigg)\Bab\bigg)
\nonumber \\ &&+{1\over (y+z)^2}2\bigg(\Bab-\Babc\bigg)
%\nonumber \\ &&
+{1\over (y+z)}2\e\bigg(\Babc-2\Bab\bigg)
\nonumber \\ &&+{y\over (1-z)^2}(1-\e)\bigg(\Babc-\Bbc\bigg)
\nonumber \\ &&+{y\over (1-z)}\bigg((3-4\e-\e^2+2\e^3)\Babc+(-3+5\e-2\e^3)\Bbc\bigg)
\nonumber \\ &&+{1\over (1-z)}(2+\e-5\e^2-2\e^3)\bigg(\Bbc-\Babc\bigg)
\nonumber \\ &&+(2-7\e+2\e^2+3\e^3)\Bbc
+(-4+10\e-4\e^2)\Bab
\nonumber \\ &&
+\Boxxz(1-2\e)\bigg((8-4\e)-{(1-y)\over z}4(1-\e)
\nonumber \\
&&\hspace{2cm}
+(y+z)(-4+4\e-6\e^2-2\e^3)
+{y^2\over z}(-2+4\e-2\e^2)\bigg).
\end{eqnarray}
Explicit formulae for the bubble and box integrals are given in the appendix.
The square of the Born amplitude is given in Eq.~(\ref{eq:T2}).

\subsection{The finite part}

The finite remainders of the one- and two-loop contributions to 
${\cal T}^{(6)}$ can be decomposed according to their colour structure and 
to their dependence on the number of quark flavours $N_F$. 
In the 
two-loop contribution, one finds moreover a term proportional to 
the charge-weighted sum of the quark flavours $N_{F,\gamma}$; this 
equals, in the case of purely electromagnetic interactions:
\begin{equation}
 N_{F,\gamma} = \frac{\left(\sum_q e_q\right)^2}{\sum_q e_q^2}\; .
\end{equation}
This term originates from diagrams containing a closed quark loop
coupling to the virtual photon and which first appear at the two-loop level. 

The tree-level combination of invariants 
\begin{equation}
T = \frac{y}{z}+\frac{z}{y}+\frac{2}{yz}-\frac{2}{y}-\frac{2}{z}
\end{equation}
frequently occurs in the finite part. We therefore extracted this 
combination by expressing $1/(yz)$ by $T$ according to the above equation.

\subsubsection{Two-loop contribution to ${\cal T}^{(6)}$}

The finite remainder of the interference of the two-loop amplitude 
with the tree-level amplitude is decomposed as
\begin{eqnarray}
\Finite^{(2 \times 0)}(x,y,z) &=& V\bigg[ 
N^2  \left(A_{20}(y,z) + A_{20}(z,y)\right)
 + \left(B_{20}(y,z)+B_{20}(z,y)\right) \nonumber \\ &&
+\frac{1}{N^2} \left( C_{20}(y,z) + C_{20}(z,y) \right) 
  + N N_F \left( D_{20} (y,z) + D_{20}(z,y) \right) \nonumber \\ &&
+\frac{N_F}{N} \left(E_{20}(y,z) + E_{20}(z,y)\right)
 +N_F^2 \left(F_{20}(y,z) + F_{20}(z,y)\right) \nonumber \\ &&
+ N_{F,\gamma} \left(\frac{4}{N}-N\right)
 \left(G_{20}(y,z) + G_{20}(z,y)\right) \bigg]\;,
\end{eqnarray} 
with


\subsubsection{One-loop contribution to ${\cal T}^{(6)}$}

The finite remainder of the self-interference of the one-loop amplitude 
is decomposed as
\begin{eqnarray}
\Finite^{(1 \times 1)}(x,y,z) &=& V\bigg[ 
N^2  \left(A_{11}(y,z) + A_{11}(z,y)\right)
 + \left(B_{11}(y,z)+B_{11}(z,y)\right) \nonumber \\ &&
+\frac{1}{N^2} \left( C_{11}(y,z) + C_{11}(z,y) \right) 
  + N N_F \left( D_{11}(y,z) + D_{11}(z,y) \right) \nonumber \\ &&
+\frac{N_F}{N} \left(E_{11}(y,z) + E_{11}(z,y)\right)
 +N_F^2 \left(F_{11}(y,z) + F_{11}(z,y)\right) \bigg]
\end{eqnarray} 
with
%

\section{Conclusions and Outlook}
\label{sec:conc}
\setcounter{equation}{0}

In this paper, we have presented analytic formulae for the two-loop virtual
corrections to the  process $\gamma^*\to q\bar q g$, which arise from the
interference of  the two-loop with the tree amplitude
and from the self-interference of the one-loop amplitude. Together with the 
contribution from the self-interference of the one-loop  amplitudes for
$\gamma^*\to ggg$~\cite{vg1,vg2}, these form the  full ${\cal
O}(\alpha_s^3)$ corrections to the three-parton subprocess  contribution to $e^+e^-
\to 3$~jets at NNLO. 

Similar results can in principle be obtained for $(2+1)$-jet production in deep
inelastic $ep$ scattering or  $(V+1)$-jet production in hadron--hadron collisions.
However, the complexity of the cut structure of the non-planar graphs together
with the rather different domains of convergence of the one- and
two-dimensional harmonic polylogarithms makes this a non-trivial task, and we
defer this to a later paper.

It must also be kept in mind that these virtual corrections form only part of a
full NNLO  calculation, which also has to include the
one-loop corrections  to $\gamma^*\to
4$~partons~\cite{onel4p1,onel4p2,onel4p3,onel4p4} where one of the
partons becomes collinear or  soft, as well as  tree-level processes
$\gamma^*\to 5$~partons~\cite{tree5p1,tree5p2,tree5p3} with two soft or collinear partons. Only
after summing all these contributions (and including  terms from the
renormalization of parton distributions  for processes with partons in the
initial state), do the divergent terms cancel  among one another.   The
factorization properties of both the one-loop, one-unresolved-parton
contribution~\cite{onel1,onel2,onel3,onel4,onel5,onel6} and the tree-level, 
two-unresolved-parton contributions~\cite{twot1,twot2,twot3,twot4} have been studied,  but a systematic
procedure for isolating the infrared singularities has not been established. 
Although this is still an open and highly non-trivial issue, significant
progress is anticipated in the near future. 

The remaining finite terms must then be combined  into a numerical program
implementing the  experimental definition of jet observables and event-shape
variables.  A first calculation involving the above features  was presented for
the case of  photon-plus-one-jet final  states in electron--positron
annihilation in~\cite{ggamma0,ggamma}, 
thus demonstrating the feasibility of this type
of calculations. A prerequisite for such a numerical program is a stable and
efficient next-to-leading order
four-jet program, where the infrared singularities 
for the one-loop $\gamma^*\to 4$~partons are combined with the tree-level 
$\gamma^* \to 5$~parton with one parton unresolved.   Four such programs
currently exist \cite{menloparc,debrecen,eerad2,mercutio}, each of which
could be used as a starting point for a  full ${\cal O}(\alpha_s^3)$
NNLO three-jet program.

\section*{Acknowledgements} 
TG and ER would like to thank Pierpaolo Mastrolia for numerous discussions 
on the automatic derivation of symbolic reduction equations. 
This work was supported in part by the EU Fourth Framework Programme 
``Training and Mobility of Researchers'', 
network ``Quantum Chromodynamics and the Deep Structure of 
Elementary Particles'', contract FMRX-CT98-0194 (DG 12-MIHT). 
 LWG thanks the UK Particle Physics and Astronomy Research Council for a
 research studentship.
AK thanks the University of Durham for a Research Studentship.
TG and EWNG would like to acknowledge the 
hospitality of the Workshop ``Physics at TeV colliders'' (Les Houches, France, 
May 2001), where this work was initiated. 

\begin{appendix}
\renewcommand{\theequation}{\mbox{\Alph{section}.\arabic{equation}}}

\section{One-loop master integrals}
\setcounter{equation}{0}
\label{app:master_int}

In this appendix, we list the expansions for the one-loop 
master integrals appearing in 
$\braket{\cm^{(0)}}{\cm^{(1)}}$ and $\braket{\cm^{(1)}}{\cm^{(1)}}$. 
These squared amplitudes can be 
expressed in terms of only two master integrals evaluated at $d=4-2\e$,
\begin{eqnarray}
{\rm Bub}(p^2) &=&\bubbleLO{p} \nonumber \\
&=& \int \frac{\d^d k}{(2\pi)^d}\, 
\frac{1}{k^2(k-p)^2},\\ 
{\rm Box}(s_{23},s_{13},s_{123})&=&
\boxLO  \nonumber \\
&=&\int \frac{\d^d k}{(2\pi)^d} \frac{1}{k^2 (k-p_2)^2
  (k-p_2-p_3)^2 (k-p_1-p_2-p_3)^2}\; . 
\end{eqnarray}
Note that in Eqs.~(\ref{eq:f1}) and (\ref{eq:f2}) we have written the one-loop
functions $f_1$ and $f_2$ in terms of the one-loop box integral in $d=6-2\e$. 
This is straightforwardly related to the box integral in $d=4-2\e$ dimensions
by
\begin{eqnarray}
{\rm Box}^{6}(s_{23},s_{13},s_{123})
&=&
-{s_{13}s_{23}\over 2(d-3)s_{12}}
{\rm Box}(s_{23},s_{13},s_{123})\nonumber \\
&&-{2\over s_{12}(d-4)}
\left({\rm Bub}(s_{13})+{\rm Bub}(s_{23})-{\rm Bub}(s_{123})\right)\;.
\end{eqnarray}

Closed expressions for these integrals for symbolic $d$ 
in terms of $\Gamma$-functions 
and the $\,_2F_1$ hypergeometric function 
have been known for a long 
time (see e.g.\ \cite{gg,gr}). The bubble integral reads
\begin{equation}
{\rm Bub}(s_{12}) = i\frac{(4\pi)^\e}{16\pi^2} \frac{ \Gamma (1+\e)
    \Gamma^2 (1-\e)}{ \Gamma (1-2\e)} \left(-s_{12}\right)^{-\e} 
\frac{1}{\e(1-2\e)} \;.
\end{equation}
In the present context, an expansion of the box
integral to the second order in $\e=(4-d)/2$ is required. 
\begin{equation}
{\rm Box}(s_{23},s_{13},s_{123}) = i\frac{(4\pi)^\e}{16\pi^2} \frac{ \Gamma (1+\e)
    \Gamma^2 (1-\e)}{ \Gamma (1-2\e)} \left(-s_{123}\right)^{-2-\e} 
\frac{1}{yz} \sum_{i=-2}^2 \frac{l_{4.1,i}
\left(\frac{\sac}{\sabc},\frac{\sbc}{\sabc}\right)}{\e^i}\; + 
{\cal O}(\e^3) ,
\end{equation}
with
\begin{eqnarray}
l_{4.1,2}(y,z) &=& 2  \ , \\
l_{4.1,1}(y,z) &=&  
          - 2 \H(0;z)
          - 2 \G(0;y)  \ , \\
l_{4.1,0}(y,z) &=&  
            2 \H(0;z) \G(0;y)
          + 2 \H(0,0;z)
          + 2 \H(1,0;z)
          + 2 \G(0,0;y)
          - 2 \G(1,0;y)
          + \frac{\pi^2}{3}
 \ , \\
l_{4.1,-1}(y,z) &=&  
            2 \H(0;z) \G(1 - z,0;y)
          - 2 \H(0;z) \G(0,0;y)
          - 2 \H(0,0;z) \G(0;y)
          - 2 \H(0,0,0;z) \nonumber \\ &&
          - 2 \H(0,1,0;z)
          + 2 \H(1,0;z) \G(1 - z;y)
          - 2 \H(1,0;z) \G(0;y)
          - 2 \H(1,0,0;z)\nonumber \\ &&
          - 2 \H(1,1,0;z)
          - 2 \G(1 - z,1,0;y)
          - 2 \G(0,0,0;y)
          + 2 \G(0,1,0;y)
          + 2 \G(1,0,0;y)
          )\nonumber \\ &&
       + \frac{\pi^2}{3}\left[
          -   \H(0;z)
          -   \H(1;z)
          +   \G(1 - z;y)
          -   \G(0;y)
          \right]
 \ , \\
l_{4.1,-2}(y,z) &=&  
            2 \H(0;z) \G(1 - z,1 - z,0;y)
          - 2 \H(0;z) \G(1 - z,0,0;y)
          - 2 \H(0;z) \G(0,1 - z,0;y)\nonumber \\ &&
          + 2 \H(0;z) \G(0,0,0;y)
          - 2 \H(0,0;z) \G(1 - z,0;y)
          + 2 \H(0,0;z) \G(0,0;y)\nonumber \\ &&
          + 2 \H(0,0,0;z) \G(0;y)
          + 2 \H(0,0,0,0;z)
          + 2 \H(0,0,1,0;z)
          - 2 \H(0,1,0;z) \G(1 - z;y)\nonumber \\ &&
          + 2 \H(0,1,0;z) \G(0;y)
          + 2 \H(0,1,0,0;z)
          + 2 \H(0,1,1,0;z)
          + 2 \H(1,0;z) \G(1 - z,1 - z;y)\nonumber \\ &&
          - 2 \H(1,0;z) \G(1 - z,0;y)
          - 2 \H(1,0;z) \G(0,1 - z;y)
          + 2 \H(1,0;z) \G(0,0;y)\nonumber \\ &&
          - 2 \H(1,0,0;z) \G(1 - z;y)
          + 2 \H(1,0,0;z) \G(0;y)
          + 2 \H(1,0,0,0;z)
          + 2 \H(1,0,1,0;z)\nonumber \\ &&
          - 2 \H(1,1,0;z) \G(1 - z;y)
          + 2 \H(1,1,0;z) \G(0;y)
          + 2 \H(1,1,0,0;z)
          + 2 \H(1,1,1,0;z)\nonumber \\ &&
          - 2 \G(1 - z,1 - z,1,0;y)
          + 2 \G(1 - z,0,1,0;y)
          + 2 \G(1 - z,1,0,0;y)
          + 2 \G(0,1 - z,1,0;y)\nonumber \\ &&
          + 2 \G(0,0,0,0;y)
          - 2 \G(0,0,1,0;y)
          - 2 \G(0,1,0,0;y)
          - 2 \G(1,0,0,0;y)          
       + \frac{7\pi^4}{180}\nonumber \\ &&
       + \frac{\pi^2}{3}   \big[
          -  \H(0;z) \G(1 - z;y)
          +  \H(0;z) \G(0;y)
          +  \H(0,0;z)
          +  \H(0,1;z)
          -  \H(1;z) \G(1 - z;y)\nonumber \\ &&
          +  \H(1;z) \G(0;y)
          +  \H(1,0;z)
          +  \H(1,1;z)
          +  \G(1 - z,1 - z;y)
          -  \G(1 - z,0;y)\nonumber \\ &&
          -  \G(0,1 - z;y)
          +  \G(0,0;y)
          \big]\;.
\end{eqnarray}

\section{Harmonic polylogarithms}
\setcounter{equation}{0}
\label{app:hpl}
The generalized polylogarithms ${\rm S}_{n,p}(x)$ 
of Nielsen~\cite{nielsen} turn out to be insufficient for the computation of 
multi-scale integrals beyond one loop. To overcome this limitation, 
one has to extend generalized polylogarithms to harmonic 
polylogarithms~\cite{hpl,mi}.

Harmonic polylogarithms are obtained by the repeated integration of 
rational factors. If these rational factors contain, besides the 
integration variable, only constants, the resulting 
functions are one-dimensional harmonic polylogarithms 
(or simply harmonic polylogarithms, HPLs)\cite{hpl,grnum1}. 
If the rational 
factors depend on a further variable, one obtains 
two-dimensional harmonic polylogarithms (2dHPLs)~\cite{mi,grnum2}.
In the following, we define both classes of functions, and summarize
their properties. 

\subsection{One-dimensional harmonic polylogarithms}

The HPLs, introduced in \cite{hpl}, are 
one-variable functions $ \H(\vec{a};x) $ depending, besides the argument 
$x$, on a set of indices, grouped for convenience into the vector 
$\vec{a}$, whose components can take one of the three values $(1,0,-1)$ 
and whose number is the weight $w$ of the HPL. More explicitly, the three 
HPLs  with $w=1$ are defined as 
\begin{eqnarray} 
  \H(1;x) &=& \int_0^x \frac{\d x'}{1-x'} = - \ln(1-x) \ , \nonumber\\ 
  \H(0;x) &=& \ln x \ ,          \nonumber\\ 
  \H(-1;x) &=& \int_0^x \frac{\d x'}{1+x'} = \ln(1+x) \ ; 
\label{eq:defineh1} 
\end{eqnarray} 
their derivatives can be written as 
\begin{equation} 
  \frac{\d }{\d x} \H(a;x) = \f(a;x) \ , \hskip 1cm a=1,0,-1 \;,
\label{eq:derive1} 
\end{equation} 
where the 3 rational fractions $\f(a;x)$ are given by 
\begin{eqnarray} 
   \f(1;x) &=& \frac{1}{1-x} \ , \nonumber\\ 
   \f(0;x) &=& \frac{1}{x} \ , \nonumber\\ 
   \f(-1;x) &=& \frac{1}{1+x} \ . 
\label{eq:definef} 
\end{eqnarray} 
For weight $w$ larger than 1, write $ \vec{a} = (a, \vec b) $, where 
$a$ is the leftmost component of  $ \vec{a} $ and $\vec b $ stands 
for the vector of the remaining $(w-1)$ components. The harmonic 
polylogarithms of weight $w$ are then defined as follows: 
if all the $w$ components of $\vec a$ take the value 0, $\vec a$ is said 
to take the value $\vec 0_w$ and 
\begin{equation} 
\H(\vec{0}_w;x) = \frac{1}{w!} \ln^w{x} \ , 
\label{eq:defh0} 
\end{equation} 
while, if $\vec{a} \neq \vec{0}_w$,
\begin{equation} 
\H(\vec{a};x) = \int_0^x \d x' \ \f(a;x') \ \H(\vec{b};x') \ . 
\label{eq:defn0} 
\end{equation} 
In any case the derivatives can be written in the compact form 
\begin{equation} 
\frac{\d }{\d x} \H(\vec{a};x) = \f(a;x) \H(\vec{b};x) \ , 
\label{eq:derive} 
\end{equation} 
where, again, $a$ is the leftmost component of $ \vec a $ and 
$ \vec b $ stands for the remaining $(w-1)$ components. 
\par 
It is immediate to see, from the very definition Eq.\ (\ref{eq:defn0}), that 
there are $3^w$ HPLs  of weight $w$, and that they are linearly 
independent. The HPLs  are generalizations of Nielsen's 
polylogarithms~\cite{nielsen}. The function $ \S_{n,p}(x) $, in 
Nielsen's notation, is equal to the HPL whose first $n$ indices are all 
equal to 0 and the remaining $p$ indices all equal to 1:
\begin{equation}
   \S_{n,p}(x) = \H(\vec{0}_{n},\vec{1}_p;x) \; ; 
\end{equation} 
in particular the Euler polylogarithms $ \Li_n(x) = \S_{n-1,1}(x) $ 
correspond to 
\begin{equation}
   \Li_n(x) = \H(\vec{0}_{n-1},1;x) \; .
\end{equation} 

As shown in \cite{hpl}, the product of two HPLs  of a same argument $x$ 
and weights $p, q$ can be expressed as a combination of HPLs  of that 
argument and weight $r=p+q$, according to the product identity 
\begin{eqnarray} 
 \H(\vec{p};x)\H(\vec{q};x) & = & 
  \sum_{\vec{r} = \vec{p}\uplus \vec{q}} \H(\vec{r};x) \; , 
\label{eq:halgebra} \end{eqnarray} 
where $\vec p, \vec q$ stand for the $p$ and $q$ components of the indices 
of the two HPLs, while $\vec{p}\uplus \vec{q}$ represents all mergers of 
$\vec{p}$ and $\vec{q}$ into the vector $\vec{r}$ with $r$ components, 
in which the relative orders of the elements of $\vec{p}$ and $\vec{q}$ 
are preserved. 
\par 
The explicit formulae relevant up to weight 4 are 
\begin{equation} 
   \H(a;x) \; \H(b;x) =  \H(a,b;x) + \H(b,a;x) \ , 
\label{eq:alg0} 
\end{equation} 
\begin{eqnarray} 
   \H(a;x) \; \H(b,c;x) &=&  \H(a,b,c;x) + \H(b,a,c;x) + \H(b,c,a;x) 
                             \ , \nonumber\\ 
   \H(a;x) \; \H(b,c,d;x) &=&  \H(a,b,c,d;x) + \H(b,a,c,d;x) + \H(b,c,a,d;x) 
                              + \H(b,c,d,a;x) \ , 
\label{eq:alg1} 
\end{eqnarray} 
and 
\begin{eqnarray} 
  \H(a,b;x) \; \H(c,d;x) &=&  \H(a,b,c,d;x) + \H(a,c,b,d;x) + \H(a,c,d,b;x) 
                              \nonumber\\
                         &+& \H(c,a,b,d;x)  + \H(c,a,d,b;x) + \H(c,d,a,b;x) 
                             \ , 
\label{eq:alg2} 
\end{eqnarray} 
where $a,b,c,d$ are indices taking any of the values $(1,0,-1)$. 
The formulae can be easily verified, one at a time, 
by observing that they are true at some specific point (such as $x=0$, 
where all the HPLs  vanish except in the otherwise trivial case in which 
all the indices are equal to 0), then taking the $x$-derivatives of the 
two sides according to Eq.\ (\ref{eq:derive}) and checking that they are equal 
(using when needed the previously established lower-weight formulae). 

Another class of identities is obtained by integrating (\ref{eq:defh0}) 
by parts. These integration-by-parts (IBP) identities read:
\begin{eqnarray}
\H(m_1,\ldots,m_q;x) &=&  \H(m_1;x)\H(m_2,\ldots,m_q;x)
                        -\H(m_2,m_1;x)\H(m_3,\ldots,m_q;x) \nonumber \\
&& + \ldots + (-1)^{q+1} \H(m_q,\ldots,m_1;x)\;.
\label{eq:ibp}
\end{eqnarray} 
These identities are not fully linearly 
independent from the product identities.

%In the context of the present calculation, only HPLs of index $(0,1)$
%appear.

A numerical implementation of the HPLs up to weight $w=4$ is 
available~\cite{grnum1}.

\subsection{Two-dimensional harmonic polylogarithms}
The 2dHPLs family 
is obtained by the repeated integration, in the variable $y$, 
of rational factors chosen, in any order, from the set $1/y$, $1/(y-1)$, 
$1/(y+z-1)$, $1/(y+z)$, where $z$ is another independent variable (hence 
the `two-dimensional' in the name). 
In full generality, let us define the rational factor 
as 
\begin{equation} 
  \g(a;y) = \frac{1}{y-a} \ , 
\label{eq:gya} 
\end{equation} 
where $a$ is the {\it index}, which can depend on $z$, $a=a(z)$; 
the rational factors which we consider for the 2dHPLs are then 
\begin{eqnarray} 
  \g(0;y) &=& \frac{1}{y} \ ,       \nonumber\\ 
  \g(1;y) &=& \frac{1}{y-1} \ ,     \nonumber\\ 
  \g(1-z;y) &=& \frac{1}{y+z-1} \ , \nonumber\\ 
  \g(-z;y) &=& \frac{1}{y+z} \ . 
\label{eq:gyalist} 
\end{eqnarray} 
With the above definitions the index takes one of the values 
$0, 1, (1-z) $ and $(-z)$. \par 
Correspondingly, the 2dHPLs at weight $w=1$ (i.e.~depending, besides 
the variable $y$, on a single further argument, or {\it index}) are 
defined to be 
\begin{eqnarray} 
 \G(0;y) &=& \ln\, y  \ ,                                \nonumber\\ 
 \G(1;y) &=& \ln\, (1-y) \ ,                            \nonumber\\ 
 \G(1-z;y) &=& \ln\left( 1 - \frac{y}{1-z} \right) \ , \nonumber\\ 
 \G(-z;y) &=& \ln\left( 1 + \frac{y}{z} \right) \ . 
\label{eq:w1list} 
\end{eqnarray} 
The 2dHPLs of weight $w$ larger than 1 depend on a set of $w$ indices, which 
can be grouped into a 
$w$-dimensional {\it vector} of indices $\vec{a}$. By writing the vector 
as $ \vec{a} = (a, \vec b)$, where $a$ is the leftmost component of 
$ \vec{a} $ and $\vec b $ stands for the vector of the remaining $(w-1)$ 
components, the 2dHPLs are then defined as follows: if all the $w$ 
components of $\vec a$ take the value 0, $\vec a$ is written as $\vec 0_w$ and 
\begin{equation} 
  \G(\vec{0}_w;y) = \frac{1}{w!} \ln^w{y} \ , 
\label{eq:defg0} 
\end{equation} 
while, if $\vec{a} \neq \vec{0}_w$, 
\begin{equation} 
  \G(\vec{a};y) = \int_0^y \d y' \ \g(a;y') \ \G(\vec{b};y') \ . 
\label{eq:defx0} 
\end{equation} 
In any case the derivatives can be written in the compact form 
\begin{equation} 
\frac{\d }{\d y} \G(\vec{a};y) = \g(a;y) \G(\vec{b};y) \ , 
\label{eq:deriveG} 
\end{equation} 
where, again, $a$ is the leftmost component of $ \vec a $ and 
$ \vec b $ stands for the remaining $(w-1)$ components. 

It should be noted that the notation for the 2dHPLs employed here is the
notation of~\cite{grnum2}, which is different 
from the original definition proposed in~\cite{mi}. Detailed
conversion rules between different notations, as well as relations to 
similar functions in the mathematical literature (hyperlogarithms and 
multiple polylogarithms) can be found in the appendix of~\cite{grnum2}. 

Algebra and reduction equations of the 2dHPLs are the same as for the 
ordinary HPLs.
The product of two 2dHPLs  of a same argument $y$ 
and weights $p, q$ can be expressed as a combination of 2dHPLs  of that 
argument and weight $r=p+q$, according to the product identity 
\begin{eqnarray} 
 \G(\vec{p};x)\G(\vec{q};x) & = & 
  \sum_{\vec{r} = \vec{p}\uplus \vec{q}} \G(\vec{r};x) \; , 
\end{eqnarray} 
where $\vec p, \vec q$ stand for the $p$ and $q$ components of the indices 
of the two 2dHPLs, while $\vec{p}\uplus \vec{q}$ represents all possible 
mergers of $\vec{p}$ and $\vec{q}$ into the vector $\vec{r}$ with $r$ 
components, in which the relative orders of the elements of $\vec{p}$ 
and $\vec{q}$ are preserved. The explicit product identities up to 
weight $w=4$ are identical to those for the HPLs 
(\ref{eq:alg0})--(\ref{eq:alg2}), with all $\H$ replaced by $\G$. 

The integration-by-parts identities read:
\begin{eqnarray}
\G(m_1,\ldots,m_q;x) &=&  \G(m_1;x)\G(m_2,\ldots,m_q;x)
                        -\G(m_2,m_1;x)\G(m_3,\ldots,m_q;x) \nonumber \\
&& + \ldots + (-1)^{q+1} \G(m_q,\ldots,m_1;x)\;.
\end{eqnarray} 

A numerical implementation of the 2dHPLs up to weight $w=4$ is 
available~\cite{grnum2}.

\end{appendix}


\begin{thebibliography}{99}

\bibitem{petra}
TASSO collaboration, D.P.~Barber et al.,
  Phys.~Rev.~Lett.~{\bf 43} (1979) 830;\\
P.~S\"oding, B.~Wiik, G.~Wolf and S.L.~Wu, Talks given at Award Ceremony 
of the 1995 EPS High Energy and Particle Physics Prize, Proceedings of 
the {\it EPS High Energy Physics Conference}, Brussels, 1995, (World
Scientific), p.~3.
%%CITATION = PRLTA,43,830;%%

\bibitem{ellis}
J.~Ellis, M.K.~Gaillard and G.G.~Ross, Nucl.~Phys.~{\bf B111} (1976) 253;
{\bf B130} (1977) 516(E).
%%CITATION = NUPHA,B111,253;%%

\bibitem{bethke}
S.~Bethke,
%``Determination of the QCD coupling alpha(s),''
J.\ Phys.\ {\bf G26} (2000) R27
[arXiv:hep-ex/0004021].
%%CITATION = HEP-EX 0004021;%%

\bibitem{ert1}
R.K.~Ellis, D.A.~Ross and A.E.~Terrano, Nucl.~Phys.~{\bf B178} (1981) 
421.
%%CITATION = NUPHA,B178,421;%%

\bibitem{ert2}
K.~Fabricius, I.~Schmitt, G.~Kramer and G.~Schierholz, Z.~Phys.~{\bf
  C11} (1981) 315.
%%CITATION = ZEPYA,C11,315;%%

\bibitem{kn}
Z.~Kunszt and P.~Nason, in {\it Z Physics at LEP 1}, CERN Yellow Report 
89-08, Vol.~1, p.~373.

\bibitem{gg}
W.T.~Giele and E.W.N.~Glover,
%``Higher order corrections to jet cross-sections in e+ e- annihilation,''
Phys.\ Rev.\  {\bf D46} (1992) 1980.
%%CITATION = PHRVA,D46,1980;%%

\bibitem{cs}
S.~Catani and M.H.~Seymour,
%``A general algorithm for calculating jet cross sections in NLO QCD,''
Nucl.\ Phys.\  {\bf B485} (1997) 291;
{\bf B510} (1997) 503(E) [arXiv:hep-ph/9605323].
%%CITATION = HEP-PH 9605323;%%

\bibitem{tesla}
R.D.\ Heuer, D.J.\ Miller, F.\ Richard and P.M.\ Zerwas (Eds.),
``TESLA Technical Design Report Part III: Physics at an $e^+e^-$ Linear 
Collider'', DESY-report 2001-011 [arXiv:hep-ph/0106315].
%%CITATION = HEP-PH 0106315;%%

\bibitem{kunszt}
Z.~Kunszt (ed.), Proceedings of the Workshop on ``New Techniques for 
Calculating Higher Order QCD Corrections'', Z\"urich, 1992, ETH-TH/93-01.

\bibitem{tree5p1}
K.~Hagiwara and D.~Zeppenfeld, Nucl.~Phys.~{\bf B313} (1989) 560.
%%CITATION = NUPHA,B313,560;%%

\bibitem{tree5p2}
F.A.~Berends, W.T.~Giele and  H.~Kuijf,  Nucl.~Phys.~{\bf B321} (1989) 39. 
%%CITATION = NUPHA,B321,39;%%

\bibitem{tree5p3}
N.K.~Falck, D.~Graudenz and G.~Kramer, Nucl.~Phys.~{\bf B328} (1989) 317.
%%CITATION = NUPHA,B328,317;%%

\bibitem{onel4p1}
Z.~Bern, L.J.~Dixon, D.A.~Kosower and S.~Weinzierl,
%``One-loop amplitudes for e+ e- $\to$ anti-q q anti-Q Q,''
Nucl.\ Phys.\  {\bf B489} (1997) 3
[arXiv:hep-ph/9610370].
%%CITATION = HEP-PH 9610370;%%

\bibitem{onel4p2}
Z.~Bern, L.J.~Dixon and D.A.~Kosower,
%``One-loop amplitudes for e+ e- to four partons,''
Nucl.\ Phys.\  {\bf B513} (1998) 3
[arXiv:hep-ph/9708239].
%%CITATION = HEP-PH 9708239;%%

\bibitem{onel4p3}
E.W.N.~Glover and D.J.~Miller,
%``The one-loop QCD corrections for gamma* $\to$ Q anti-Q q anti-q,''
Phys.\ Lett.\  {\bf B396} (1997) 257
[arXiv:hep-ph/9609474].
%%CITATION = HEP-PH 9609474;%%

\bibitem{onel4p4}
J.M.~Campbell, E.W.N.~Glover and D.J.~Miller,
%``The one-loop QCD corrections for gamma* $\to$ q anti-q g g,''
Phys.\ Lett.\  {\bf B409} (1997) 503
[arXiv:hep-ph/9706297].
%%CITATION = HEP-PH 9706297;%%

%\bibitem{onel4p5}
%Z.~Nagy and Z.~Trocsanyi,
%``Group independent color decomposition of next-to-leading order matrix  elements for e+ e- $\to$ four partons,''
%Phys.\ Lett.\  {\bf B414} (1997) 187
%[arXiv:hep-ph/9708342].
%%CITATION = HEP-PH 9708342;%%


\bibitem{hv}
G.\ 't Hooft and M.\ Veltman, Nucl.\ Phys.\ {\bf B44} (1972) 189.
%%CITATION = NUPHA,B44,189;%%

\bibitem{chet1}
F.V.\ Tkachov, Phys.\ Lett.\ {\bf 100B} (1981) 65.
%%CITATION = PHLTA,B100,65;%%
\bibitem{chet2}
K.G.\ Chetyrkin and F.V.\ Tkachov, Nucl.\ Phys.\ {\bf B192} (1981) 159.
%%CITATION = NUPHA,B192,159;%%

\bibitem{gr} 
T.\ Gehrmann and E.\ Remiddi, Nucl.\ Phys.\
{\bf B580} (2000) 485 [arXiv:hep-ph/9912329].
%%CITATION = HEP-PH 9912329;%%


\bibitem{m1}
Z.~Bern, L.~Dixon and A.~Ghinculov, Phys.\ Rev.\ {\bf D63} (2001) 053007
[arXiv:hep-ph/0010075].
%%CITATION = HEP-PH 0010075;%%

\bibitem{m2}
C.\ Anastasiou, E.W.N.~Glover, C.\ Oleari and M.E.\ Tejeda-Yeomans,
Nucl.\ Phys.\ {\bf B601} (2001) 318 [arXiv:hep-ph/0010212]; 
{\bf B601} (2001) 347 [arXiv:hep-ph/0011094]. 
%%CITATION = HEP-PH 0010212;%%
%%CITATION = HEP-PH 0011094;%%

\bibitem{m3}
C.\ Anastasiou, E.W.N.~Glover, C.\ Oleari and M.E.\ Tejeda-Yeomans,
Nucl.\ Phys.\ 
{\bf B605} (2001) 486 [arXiv:hep-ph/0101304].
%%CITATION = HEP-PH 0101304;%%

\bibitem{m4}
E.W.N.~Glover, C.~Oleari and M.E.~Tejeda-Yeomans,
%``Two-loop QCD corrections to gluon gluon scattering,''
Nucl.\ Phys.\ {\bf B605} (2001) 467 [arXiv:hep-ph/0102201].
%%CITATION = HEP-PH 0102201;%%


\bibitem{m5}
Z.~Bern, A.~De Freitas and L.J.~Dixon,
%``Two-loop amplitudes for gluon fusion into two photons,''
JHEP {\bf 0109} (2001) 037 [arXiv:hep-ph/0109078].
%%CITATION = HEP-PH 0109078;%%

\bibitem{m6}
Z.~Bern, A.~De Freitas, L.J.~Dixon, A.~Ghinculov and H.L.~Wong,
%``QCD and QED corrections to light-by-light scattering,''
JHEP {\bf 0111} (2001) 031
[arXiv:hep-ph/0109079].
%%CITATION = HEP-PH 0109079;%%

\bibitem{onshell1}
V.A.\ Smirnov, Phys.\ Lett.\ {\bf B460} (1999) 397 [arXiv:hep-ph/9905323].
%%CITATION = HEP-PH 9905323;%%
\bibitem{onshell2}
J.B.\ Tausk, Phys.\ Lett.\ {\bf B469} (1999) 225 [arXiv:hep-ph/9909506].
%%CITATION = HEP-PH 9909506;%%
\bibitem{onshell3}
V.A.\ Smirnov and O.L.\ Veretin,  Nucl.\ Phys.\ {\bf B566}
(2000) 469 [arXiv:hep-ph/9907385].
%%CITATION = HEP-PH 9907385;%%
\bibitem{onshell4}
C.\ Anastasiou, T.\ Gehrmann, C.\ Oleari, E.\ Remiddi and 
J.B.\ Tausk, Nucl.\ Phys.\
{\bf B580} (2000) 577 [arXiv:hep-ph/0003261].
%%CITATION = HEP-PH 0003261;%%
\bibitem{onshell5}
T.~Gehrmann and E.~Remiddi, Nucl.\ Phys.\ {\bf B} (Proc.\ Suppl.)
{\bf 89} (2000) 251 [arXiv:hep-ph/0005232].
%%CITATION = HEP-PH 0005232;%%
\bibitem{onshell6}
C.\ Anastasiou, J.B.\ Tausk and M.E.\ Tejeda-Yeomans, 
Nucl.\ Phys.\ {\bf B} (Proc.\ Suppl.) {\bf 89} (2000) 262
[arXiv:hep-ph/0005328].
%%CITATION = HEP-PH 0005328;%%




\bibitem{mi}
T.\ Gehrmann and E.\ Remiddi, Nucl.~Phys.~{\bf B601} (2001) 248
[arXiv:hep-ph/0008287];
{\bf B601} (2001) 287 [arXiv:hep-ph/0101124].
%%CITATION = HEP-PH 0008287;%%
%%CITATION = HEP-PH 0101124;%%

\bibitem{num}
T.\ Binoth and G.\ Heinrich, Nucl.\ Phys.\ {\bf B585} (2000) 741
[arXiv:hep-ph/0004013].
%%CITATION = HEP-PH 0004013;%%

\bibitem{smirnov}
V.A.~Smirnov, Phys.\ Lett.\ {\bf B491} (2000) 130 [arXiv:hep-ph/007032]; 
{\bf B500} (2001) 330 [arXiv:hep-ph/0011056].
%%CITATION = HEP-PH 0007032;%%
%%CITATION = HEP-PH 0011056;%%




\bibitem{hpl}
E.\ Remiddi and J.A.M.\ Vermaseren, Int.\ J.\ Mod.\ Phys.\ {\bf A15}
(2000) 725 [arXiv:hep-ph/9905237].
%%CITATION = HEP-PH 9905237;%%

\bibitem{nielsen}
N.~Nielsen, Nova Acta Leopoldiana (Halle) {\bf 90} (1909) 123.

\bibitem{bit}
K.S.\ K\"olbig, J.A.\ Mignaco and E.\ Remiddi, BIT {\bf 10} (1970) 38.





\bibitem{grnum1}
T.~Gehrmann and E.~Remiddi,
%``Numerical evaluation of harmonic polylogarithms,''
Comput.\ Phys.\ Commun.\ {\bf 141} (2001) 296 [arXiv:hep-ph/0107173].
%%CITATION = HEP-PH 0107173;%%

\bibitem{grnum2}
T.~Gehrmann and E.~Remiddi, CERN-TH/2001-326 [arXiv:hep-ph/0111255].
%%CITATION = HEP-PH 0111255;%%

\bibitem{dreg1}
C.G.\ Bollini and J.J.\ Giambiagi, Nuovo Cim.\ {\bf 12B} (1972) 20.
%%CITATION = NUCIA,B12,20;%%

\bibitem{dreg2}
G.M.\ Cicuta and E.\ Montaldi, Nuovo Cim.\ Lett.\ {\bf 4} (1972) 329.
%%CITATION = NCLTA,4,329;%%


\bibitem{catani}
S.\ Catani, Phys.\ Lett.\ {\bf B427} (1998) 161 [arXiv:hep-ph/9802439].
%%CITATION = HEP-PH 9802439;%%


\bibitem{vg1}
V.N.\ Baier, E.A.\ Kurayev and V.S.\ Fadin, Sov.\ J.\ Phys.\ {\bf 31} 
(1980) 364.

\bibitem{vg2}
J.J.~van der Bij and E.W.N.~Glover,
%``Z Boson Production And Decay Via Gluons,''
Nucl.\ Phys.\ B {\bf 313} (1989) 237.
%%CITATION = NUPHA,B313,237;%%

\bibitem{qgraf}
P.~Nogueira,
%``Automatic Feynman graph generation,''
J.\ Comput.\ Phys.\  {\bf 105} (1993) 279.
%%CITATION = JCTPA,105,279;%%


\bibitem{maple}
{\it MAPLE V Release 7},  Copyright 2001 
by Waterloo Maple Inc.
%

\bibitem{form2}
J.A.M.~Vermaseren, {\it Symbolic Manipulation with FORM}, Version 2,
CAN, Amsterdam, 1991.
% 
\bibitem{form3}
J.A.M.~Vermaseren, 
%``New features of FORM,''
arXiv:math-ph/0010025.
%%CITATION = MATH-PH 0010025;%%


\bibitem{pv}
G.~Passarino and M.J.~Veltman,
%``One Loop Corrections For E+ E- Annihilation Into Mu+ Mu- In The Weinberg Model,''
Nucl.\ Phys.\ B {\bf 160} (1979) 151.
%%CITATION = NUPHA,B160,151;%%



\bibitem{kl1}
R.J.\ Gonsalves, Phys.\ Rev.\ {\bf D28} (1983) 1542.
%%CITATION = PHRVA,D28,1542;%
\bibitem{kl2}
G.\ Kramer and B.\ Lampe, J.\ Math.\ Phys.\ {\bf 28} (1987) 945.
%%CITATION = JMAPA,28,945;%%
\bibitem{kl3}
 W.L.\ van Neerven, Nucl.\ Phys.\ B {\bf 268} (1986) 453.
%%CITATION = NUPHA,B268,453;%%

\bibitem{nigelpentabox}
C.~Anastasiou, E.W.N.~Glover and C.~Oleari,
%``The two-loop scalar and tensor pentabox graph with light-like legs,''
Nucl.\ Phys.\  {\bf B575} (2000) 416; {\bf B585} (2000) 763(E)
[arXiv:hep-ph/9912251].
%%CITATION = HEP-PH 9912251;%%


\bibitem{laporta}
S.~Laporta,
%``High-precision calculation of multi-loop Feynman integrals by  difference equations,''
Int.\ J.\ Mod.\ Phys.\ A {\bf 15} (2000) 5087
[arXiv:hep-ph/0102033].
%%CITATION = HEP-PH 0102033;%%


\bibitem{qff1} 
G.\ Kramer and B.\ Lampe, Z.\ Phys.\ {\bf C34} (1987) 497;
{\bf C42} (1989) 504(E).
%%CITATION = ZEPYA,C34,497;%%
\bibitem{qff2}
T.\ Matsuura and W.L.\ van Neerven, Z.\ Phys.\ {\bf C38} (1988) 623.
%%CITATION = ZEPYA,C38,623;%%
\bibitem{qff3}
T.\ Matsuura, S.C.\ van der Maarck and W.L.\ van Neerven, 
Nucl. Phys. {\bf B319} (1989) 570.
%%CITATION = NUPHA,B319,570;%%


\bibitem{onel1}
Z.\ Bern, L.J.\ Dixon, D.C.\ Dunbar and D.A.\ Kosower,
Nucl.\ Phys.\ {\bf B425} (1994) 217
[arXiv:hep-ph/9403226].
%%CITATION = HEP-PH 9403226;%%

\bibitem{onel2}
D.A.\ Kosower, Nucl.\ Phys.\ {\bf B552} (1999) 319 [arXiv:hep-ph/9901201].
%%CITATION = HEP-PH 9901201;%%

\bibitem{onel3}
D.A.~Kosower and P.~Uwer, Nucl.\ Phys.\ {\bf B563} (1999) 477
[arXiv:hep-ph/9903515].
%%CITATION = HEP-PH 9903515;%%

\bibitem{onel4}
Z.\ Bern, V.\ Del Duca and C.R.\ Schmidt, Phys.\ Lett.\ {\bf B445}
(1998) 168 [arXiv:hep-ph/9810409].
%%CITATION = HEP-PH 9810409;%%

\bibitem{onel5}
Z.\ Bern, V.\ Del Duca, W.B.\ Kilgore and C.R.\ Schmidt, Phys.\ Rev.\
{\bf D60} (1999) 116001 [arXiv:hep-ph/9903516].
%%CITATION = HEP-PH 9903516;%%

\bibitem{onel6}
S.\ Catani and M.\ Grazzini, Nucl.\ Phys.\  {\bf B591} (2000) 435
[arXiv:hep-ph/0007142].
%%CITATION = HEP-PH 0007142;%%

\bibitem{twot1}
J.M.\ Campbell and E.W.N.\ Glover, 
Nucl.\ Phys.\ {\bf B527} (1998) 264 [arXiv:hep-ph/9710255].
%%CITATION = HEP-PH 9710255;%%

\bibitem{twot2}
S.\ Catani and M.\ Grazzini, Phys.\ Lett.\ {\bf B446} (1999) 143
[arXiv:hep-ph/9810389];
%%CITATION = HEP-PH 9810389;%% 
Nucl. Phys. {\bf B570} (2000) 287 [arXiv:hep-ph/9908523].
%%CITATION = HEP-PH 9908523;%%

\bibitem{twot3}
F.A.\ Berends and W.T.\ Giele, Nucl.\ Phys.\ {\bf B313} (1989) 595.
%%CITATION = NUPHA,B313,595;%%

\bibitem{twot4}
S.\ Catani, in~\cite{kunszt}.

\bibitem{ggamma0}
A.~Gehrmann-De Ridder, T.~Gehrmann and E.W.N.~Glover,
%``Radiative corrections to the photon + 1jet rate at LEP,''
Phys.\ Lett.\ B {\bf 414} (1997) 354
[arXiv:hep-ph/9705305].
%%CITATION = HEP-PH 9705305;%%

\bibitem{ggamma}
A.~Gehrmann-De Ridder and E.W.N.~Glover, Nucl.~Phys.\ {\bf B517} (1998) 269
[arXiv:hep-ph/9707224].
%%CITATION = HEP-PH 9707224;%%

\bibitem{menloparc} 
L.J.\ Dixon and A.\ Signer,
Phys.\ Rev.\ Lett.\  {\bf 78} (1997) 811
[arXiv:hep-ph/9609460];
%%CITATION = HEP-PH 9609460;%%
Phys.\ Rev.\ D {\bf 56} (1997) 4031
[arXiv:hep-ph/9706285].
%%CITATION = HEP-PH 9706285;%%

\bibitem{debrecen} 
Z.\ Nagy and Z.\ Trocsanyi,
Phys.\ Rev.\ Lett.\  {\bf 79} (1997) 3604
[arXiv:hep-ph/9707309].
%%CITATION = HEP-PH 9707309;%%

\bibitem{eerad2} 
J.M.\ Campbell, M.A.\ Cullen and E.W.N.\ Glover,
%``Four jet event shapes in electron positron annihilation,''
Eur.\ Phys.\ J.\ C {\bf 9} (1999) 245
[arXiv:hep-ph/9809429].
%%CITATION = HEP-PH 9809429;%%

\bibitem{mercutio} 
S.~Weinzierl and D.A.~Kosower,
%``{QCD} corrections to four-jet production and three-jet structure in e+ e-  annihilation,''
Phys.\ Rev.\ D {\bf 60} (1999) 054028
[arXiv:hep-ph/9901277].
%%CITATION = HEP-PH 9901277;%%


\end{thebibliography}
\end{document}